\documentclass[]{aa}  
\usepackage[varg]{txfonts}
\usepackage{natbib, twoopt}
\usepackage{graphicx}
\usepackage{amsmath}
\usepackage[psdextra,pdfencoding=auto,breaklinks=true,colorlinks=true,citecolor=blue]{hyperref}
\bibpunct{(}{)}{;}{a}{}{,}
\usepackage{color}
\definecolor{red}{rgb}{1.0,0.0,0.0}

\begin{document} 

\title{Direct imaging discovery of a super-Jovian around the young Sun-like star AF Leporis\thanks{Based on observations collected at the European Organisation for Astronomical Research in the Southern Hemisphere under ESO program 109.23AQ.001.}}
\author{Robert J. De Rosa\inst{1} \and 
        Eric L. Nielsen\inst{2} \and
        Zahed Wahhaj\inst{1} \and
        Jean-Baptiste Ruffio\inst{3} \and
        Paul G. Kalas \inst{4,5,6} \and
        Anne E. Peck\inst{2} \and
        Lea A. Hirsch\inst{7} \and
        William Roberson\inst{8}}

\institute{European Southern Observatory, Alonso de C\'{o}rdova 3107, Vitacura, Santiago, Chile \\ \email{rderosa@eso.org}
\and Department of Astronomy, New Mexico State University, P.O. Box 30001, MSC 4500, Las Cruces, NM 88003, USA
\and Center for Astrophysics and Space Science, University of California, San Diego; La Jolla, CA 92093, USA
\and Astronomy Department, University of California, Berkeley, CA 94720, USA
\and Institute of Astrophysics, FORTH, GR-71110 Heraklion, Greece
\and SETI Institute, Carl Sagan Center, 339 Bernardo Ave.,  Mountain View CA 94043, USA
\and Department of Chemical and Physical Sciences, University of Toronto, 3359 Mississauga Road, Mississauga ON L5L 1C6, Canada
\and Physics Department, Stanford University, Stanford, CA 94305, USA}

\date{Received 10 January 2023; Accepted 13 February 2023}

\abstract{Expanding the sample of directly imaged companions to nearby, young stars that are amenable to detailed astrometric and spectroscopic studies is critical for the continued development and validation of theories of their evolution and atmospheric processes.}
{The recent release of the {\it Gaia} astrometric catalog allows us to efficiently search for these elusive companions by targeting those stars that exhibit the astrometric reflex motion induced by an orbiting companion. The nearby (27\,pc), young (24\,Myr) star AF Leporis (AF Lep) was targeted because of its significant astrometric acceleration measured between the {\it Hipparcos} and {\it Gaia} astrometric catalogs, consistent with a wide-orbit planetary-mass companion detectable with high-contrast imaging.}
{We used the SPHERE instrument on the VLT to search for faint substellar companions in the immediate vicinity of AF Lep. We used observations of a nearby star interleaved with those of AF Lep to efficiently subtract the residual point spread function. This provided sensitivity to faint planetary-mass companions within $1\arcsec$ ($\sim$30\,au) of the star.}
{We detected the companion AF Lep b at a separation of 339\,mas (9\,au) from the host star, at almost the exact location predicted by the astrometric acceleration, and within the inner edge of its unresolved debris disk. The measured $K$-band contrast and the age of the star yield a model-dependent mass of between 4 and 6\,$M_{\rm Jup}$, consistent with the mass derived from an orbital fit to the absolute and relative astrometry of $4.3_{-1.2}^{+2.9}$\,$M_{\rm Jup}$. The near-infrared spectral energy distribution of the planet is consistent with an object at the L--T spectral type transition, but under-luminous with respect to field-gravity objects.}
{AF Lep b joins a growing number of substellar companions imaged around stars in the young $\beta$ Pictoris moving group. With a mass of between 3 and 7\,$M_{\rm Jup}$, it occupies a gap in this isochronal sequence between hotter, more massive companions, such as PZ~Tel~B and $\beta$~Pic~b, and the cooler 51~Eri~b, which is sufficiently cool for methane to form within its photosphere. Lying at the transition between these two classes of objects, AF Lep b will undoubtedly become a benchmark for studies of atmospheric composition and processes, as well as an anchor for models of the formation and evolution of substellar and planetary-mass companions.}


\maketitle

\section{Introduction}
High-contrast direct imaging surveys of young (1--100 Myr) nearby stars probe the occurrence rates and physical properties of Jupiter-mass planets orbiting beyond $>$3 au. To date, the general conclusion is that the formation of 2--13 $M_J$ planets with $3<a<100$ au is more likely around early-type stars with $M>1.5 M_\odot$ \citep{Nielsen:2019cb}. But with occurrence rates of wide-separation giant planets of a few percent, astrometric accelerations allow us to select potential planet hosts more efficiently compared to blind surveys. In this manuscript we present the direct detection of a 3--7 $M_J$ companion to the $\sim$1.2 $M_\odot$ star AF Leporis (\object{AF Lep}), selected due to the significant astrometric acceleration the planet induces upon the host star.

\object{AF Lep} (HD 35850, HIP 25486, HR 1817) is an F8V \citep{Gray:2006ca} star at 26.8 pc with an estimated mass of $1.29\pm0.20$\,$M_\odot$ \citep{Stassun:2019bo}. It is a likely member of the $\beta$ Pictoris moving group based on youth indicators and kinematics \citep{Zuckerman:2001go, Malo:2012gn, Ujjwal:2020dh} with age $24\pm3$\,Myr \citep{Bell:2015gw}. A query of the BANYAN~$\Sigma$ tool yields a membership probability greater than 99\% \citep{Gagne:2018jj}. \citet{Nordstrom:2004ci} classified the star as a spectroscopic binary, but \citet{ZunigaFernandez:2021if} find no significant variation. Twenty radial velocity measurements from the High Resolution Echelle Spectrometer (HIRES) at Keck spanning 11 years \citep{TalOr:2019gm} show no strong evidence of a periodic signal (see Sect.~\ref{sec:rvs}). The star is classified as an RS CVn variable (e.g., \citealp{Martinez:2022jz}), but activity indicators could easily be ascribed to the youth of the star, rather than due to interaction with a lower-mass stellar companion. The star has been targeted in previous direct imaging surveys (e.g., \citealp{Lowrance:2005ci, Brandt:2014hc, Bonavita:2016id, Galicher:2016hg, Hagan:2018ij, Nielsen:2019cb}), but no companion had been detected.

AF Lep hosts a circumstellar debris disk that was first detected as unresolved thermal infrared excess emission with the {Infrared Astronomical Satellite} and the {Infrared Space Observatory} \citep{Spangler:2001jg}. \citet{Zuckerman:2004kb} calculated $\tau$=0.2$\times10^{-4}$ for the dust fractional infrared luminosity and a blackbody dust temperature $T_{bb}=45$\,K, placing the emitting grains at $r_{bb}=55$\,au. Using additional {\it Spitzer} observations at 24, 33, and 70\,$\mu$m, \citet{Hillenbrand:2008eo} found a hotter dust temperature with $T_{bb}=85$\,K and $r_{bb}=15$\,au. This was followed by {\it Herschel} detections, from which \citet{RiviereMarichalar:2014jm} derived $T_{bb}=77.6^{+8.5}_{-15}$\,K and $r_{bb}=16.7^{+10.3}_{-7.5}$\,au using a modified blackbody model. Adding the {Wide-field Infrared Survey Explorer} detection at 22 $\mu$m to these prior data, \citet{Pawellek:2021cr} found $T_{bb}=74\pm4$\,K and $r_{bb}=19\pm8$\,au using a modified blackbody model. In summary, these various analyses of infrared photometry indicate a disk structure with a central dust depletion that serves as indirect evidence for a planetary system within a radius of $\sim$10\,au that gravitationally prevents grains from migrating inward due to Poynting-Robertson drag and filling the central disk hole. To date, the dust disk has not been spatially resolved via thermal or optical/infrared high-resolution imaging (e.g., Appendix~\ref{ap:hst}).

In this article we report on the discovery of a giant planet companion to this star, identified by the astrometric reflex motion it is inducing on the host star. We describe the target selection in Sect.~\ref{sec:sample} and discuss several aspects of the target in Sect.~\ref{sec:aflep}. The observations and data reduction are described in Sect.~\ref{sec:obs}. We report on the astrometric and spectro-photometric measurements that we used to constraint the properties of the companion in Sect.~\ref{sec:results}. We discuss the unlikely possibility that the resolved companion is an unassociated background or foreground object in Sect.~\ref{sec:alternatives}. We conclude in Sect.~\ref{sec:discussion}. An independent and contemporaneous discovery of this companion via direct imaging is reported in \citet{Mesa:2023} and \citet{Franson:2023}.

\section{Targeted imaging searches}
\label{sec:sample}
 \begin{figure}
    \centering
    \includegraphics[width=0.5\textwidth]{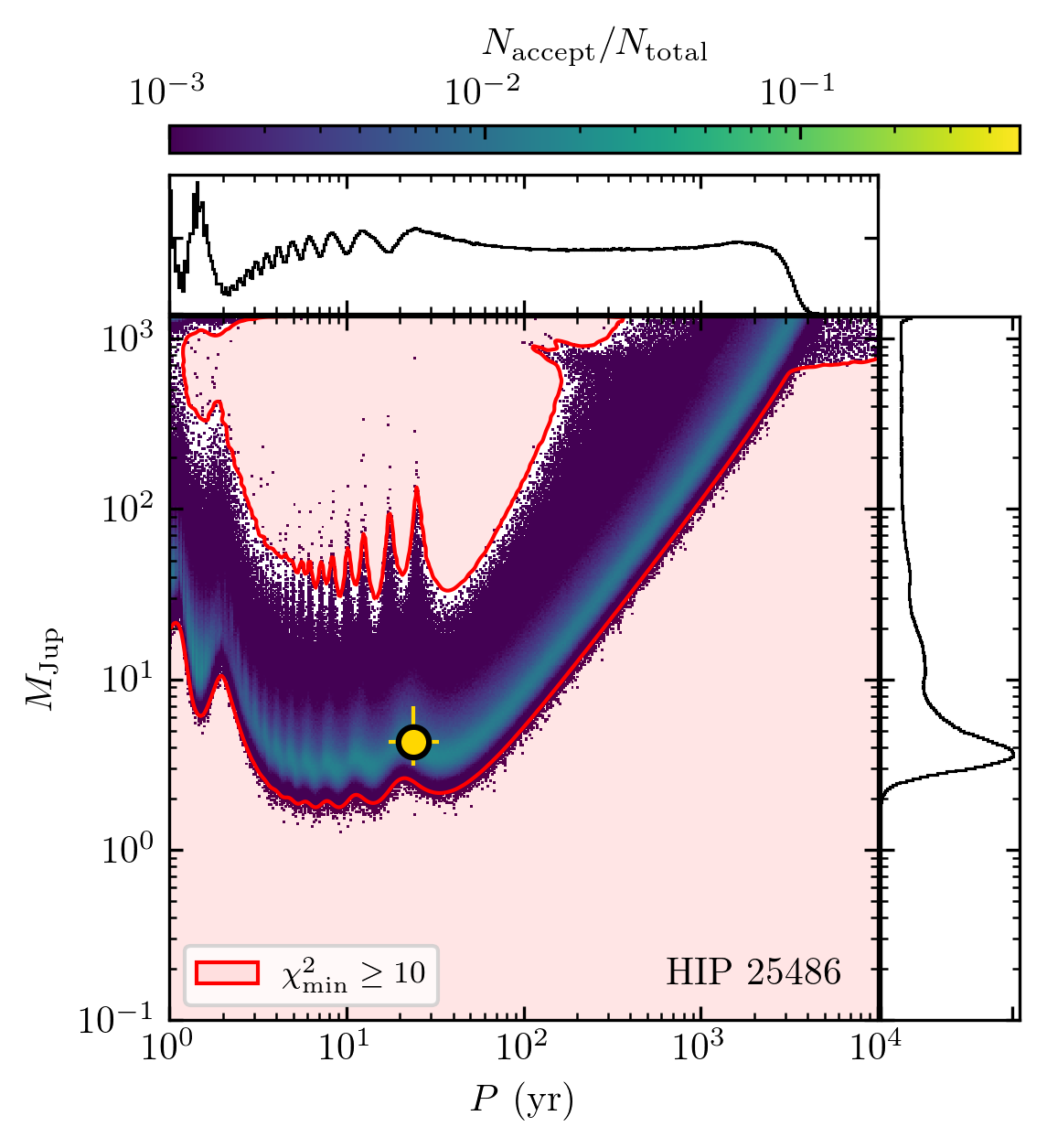}
    \caption{\label{fig:p-m}Companion masses and orbital periods consistent with the astrometric acceleration of the AF Lep photocenter reported in the HGCA. Companions in the red shaded region are excluded at a high significance. The mass and orbital period of AF Lep b estimated from the \texttt{orvara} fit described in Sect.~\ref{sec:astrometry} are indicated.}
\end{figure}
The astrometric reflex motion induced on a star by long-period substellar companions (e.g., \citealp{Brandt:2018dj,Kervella:2019bw}) can be used to dramatically increase the efficiency of direct imaging searches for brown dwarf and planetary-mass companions to nearby, young stars (e.g., \citealp{Bonavita:2022fd}). This technique leverages the precision of the {\it Hipparcos} \citep{ESA:1997ws,vanLeeuwen:2007dc} and {\it Gaia} \citep{GaiaCollaboration:2021ev} astrometric catalogs to detect the small deviations of the position and motion of the star through space as it orbits the barycenter of the system, caused by the gravitational influence of an orbiting companion. Previous detections using this technique have been limited to companions within the stellar or brown dwarf regime (e.g., \citealp{Currie:2020hq, Chilcote:2021bj, Kuzuhara:2022kc, Franson:2022vf}); however, this technique is sensitive enough to detect the signal induced by a planetary mass companion around the closest, youngest stars.

Despite the existence of large-scale direct imaging surveys for exoplanets conducted with the latest generation of high-contrast imaging instruments (e.g., \citealp{Nielsen:2019cb,Vigan:2021dc}), there is still a non-negligible discovery space around nearby, young stars. For the closest stars, orbital motion of planetary companions on $\sim$10\,au ($P\sim30$\,yr) orbits is such that companions may move over a few years from an undetectable to detectable separation, without any intervening improvement in instrumentation capabilities. There are also a handful of stars that are known to be members of nearby, young kinematic associations that have yet to be observed by any high-contrast imaging instrument. A star with a significant astrometric acceleration that falls into either of these two categories is a candidate for further high-contrast imaging observations. 

We used the Hipparcos-\textit{Gaia} Catalog of Accelerations (HGCA; \citealp{Brandt:2021cd}) to select young ($<$100\,Myr) stars that had small but statistically significant astrometric accelerations measured between the {\it Hipparcos} and {\it Gaia} catalogs. Restricting our search to stars visible from the Southern Hemisphere, we arrived at a sample of three high-priority targets whose astrometric signal was consistent with an orbiting planetary-mass companion that could be spatially resolved. AF Lep was the first star with a successfully completed observation in this program, demonstrating the significantly improved efficiency of targeted searches over blind surveys. This star has a significant astrometric acceleration. The HGCA reports a $\chi^2_{\nu}=39$ with two degrees of freedom when the astrometric measurements are fit to linear motion, strong evidence of perturbation by an orbiting companion. No evidence of a wide stellar companion to this star was found in the literature.

The ranges of companion masses and orbital periods consistent with the astrometric signal of AF Lep were estimated through a rejection sampling-based approach. We compared simulated astrometric measurements of the photocenter of a system with a massive orbiting companion to the calibrated proper motion differences given in the HGCA. We limited this analysis to circular orbits to limit the dimensionality of the problem. We generated $10^{10}$ orbits, from which $7.7\times10^6$ orbits were accepted via rejection-sampling. These accepted orbits are plotted in a two-dimensional histogram in Fig.~\ref{fig:p-m}. The accepted orbits are restricted to a narrow path through this phase space, with many areas excluded at high significance ($\chi^2_{\rm min} > 10$). From this analysis it was clear that the orbiting companion was a (1) short-period stellar or brown dwarf companion, mostly excluded by the flat radial velocity curve; (2) a long-period stellar companion, excluded in prior imaging observations \citep{Brandt:2014hc,Bonavita:2016id,Nielsen:2019cb}; or (3) a planetary-mass companion with an orbital period of $\sim$10--100 years. This third scenario could be tested given the current sensitivity of high-contrast imaging instruments.

\section{AF Lep}
\label{sec:aflep}
\subsection{Radial velocities}
\label{sec:rvs}
Close stellar companions on short-period ($<$1\,yr) orbits can bias the proper motion measurement in either of the two astrometric catalogs used to measure the proper motion anomaly reported in the HGCA, mimicking the signal induced by a long-period substellar companion. We used radial velocity measurements from Keck/HIRES \citep{TalOr:2019gm} and new measurements taken using the ARC Echelle Spectrograph (ARCES; \citealp{Wang:2003fv}) at Apache Point Observatory (APO) to search for short-period stellar companions. This search was in part motivated by the categorization of AF Lep as a spectroscopic \citep{Nordstrom:2004ci} eclipsing (e.g., \citealp{Martinez:2022jz}) binary, with a mass ratio of $q=0.715\pm0.072$. A binary companion of this mass on an edge-on orbit would induce a radial velocity semi-amplitude of between 10 and 100\,km\,s$^{-1}$ for periods between $10^0$ and $10^3$\,d, easily detectable given the radial velocity precision of the two instruments.

\begin{figure}
    \centering
    \includegraphics[width=0.5\textwidth]{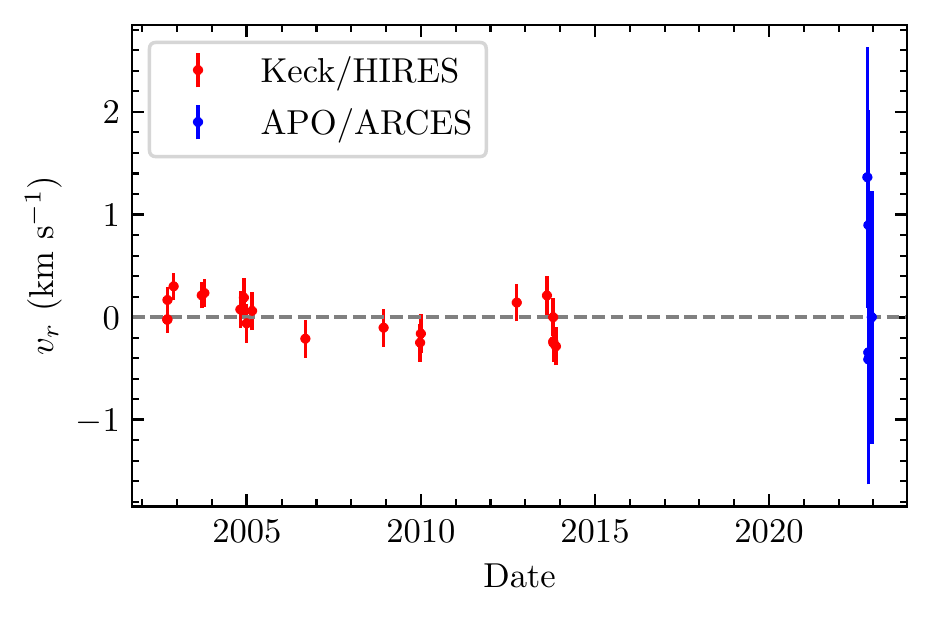}
    \caption{\label{fig:rvs}Relative radial velocity of AF Lep measured with HIRES (red) and APO (blue). The uncertainties on the HIRES observations have been inflated by approximately 100\,m\,s$^{-1}$ to better match the stellar jitter. No significant signal at the level expected for a short-period stellar companion is seen.}
\end{figure}

APO/ARCES echelle spectra in the optical were obtained for AF Lep on five epochs in late 2022, and the raw data were reduced using standard Image Reduction and Analysis Facility (IRAF) routines with a Python/PyRAF wrapper.  We clearly detect lithium absorption and calcium H\&K emission in the spectrum, consistent with literature measurements \citep{Wright:2004eb, Gray:2006ca}.  As expected for a $\sim$1 day rotation period (Sect.~\ref{sec:tess}), the absorption lines are significantly broadened.  Nevertheless, we only observe one set of lines in the spectrum.  We compute relative radial velocities from the data using custom code for ARCES spectra.  The epoch with the highest signal-to-noise ratio (S/N) is designated the template epoch, and for each remaining epoch we compute the radial velocity offset for stellar lines over multiple wavelength regions compared to the template. The process is repeated for telluric lines as well to correct for instrumental drift.  Our method reaches a precision of 100\,m\,s$^{-1}$ for narrow-lined radial velocity standards; for the rapidly rotating AF Lep, our precision is closer to $\sim$1\,km\,s$^{-1}$.  The relative radial velocities for AF Lep are given in Table~\ref{tbl:apo-rvs}.  No significant radial velocity variation is seen over the 48 days of APO/ARCES observations, consistent with the HIRES record (Fig.~\ref{fig:rvs}).

\begin{table}[h!]
\caption{Relative radial velocities of AF Lep measured with APO/ARCES.}
\label{tbl:apo-rvs}
 \centering
\begin{tabular}{ccc}
\hline\hline
JD & $v_r$ (km\,s$^{-1}$) & $\sigma_{v_r}$ (km\,s$^{-1}$)\\
\hline
2459881.77501 & $1.36$ & $1.27$ \\
2459891.74923 & $-0.41$ & $1.22$ \\
2459891.75486 & $-0.34$ & $1.25$ \\
2459892.79489 & $0.90$ & $1.12$ \\
2459928.71905 & $\equiv0$ & $1.23$ \\
\hline
\end{tabular}
\end{table}

\subsection{TESS light curve}
\label{sec:tess}
AF Lep has been observed by the Transiting Exoplanet Survey Satellite (TESS) in sectors five and six in 2018, and sector 32 in 2020. We downloaded calibrated light curves from the National Aeronautics and Space Administration (NASA) Ames Science Processing Operations Center (SPOC; \citealp{Jenkins:2016ko}). These data show a clear modulation caused by the rapid rotation of the spotted surface of the star in and out of view ($P_{\rm rot}=0.9660\pm0.0023$\,d; \citealp{Jarvinen:2015fn}). The variability is more regular in the first two periods; the data from 2020 show evidence of a longer-period variability, perhaps due to changes in the spot coverage of the star.

\begin{figure}
    \centering
    \includegraphics[width=0.5\textwidth]{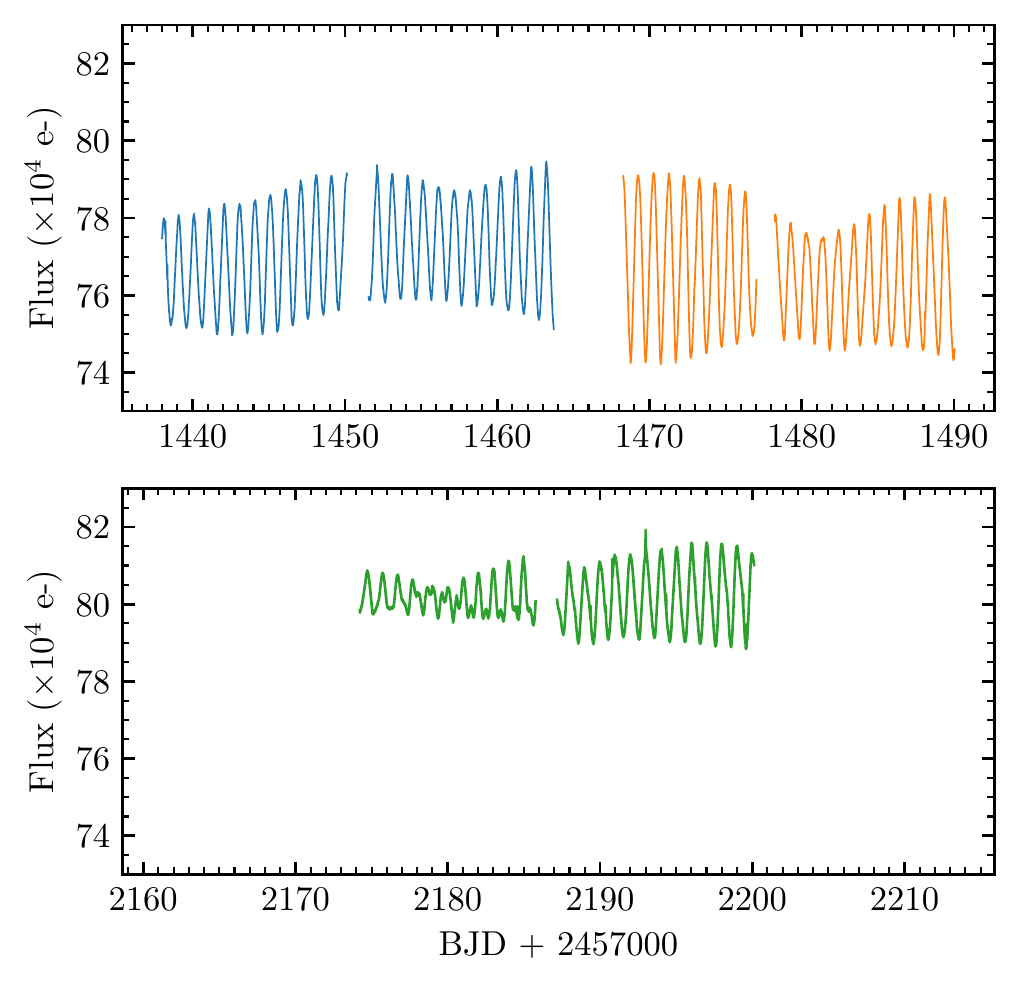}
    \caption{\label{fig:tess}{TESS} light curves of AF Lep obtained in 2018 (sectors five and six, top panel) and 2020 (sector 32, bottom panel). The $\sim$1\,d rotation period is clearly visible, although the amplitude varies between the two epochs.}
\end{figure}
We used \texttt{LightKurve} \citep{LightkurveCollaboration:2018tf} to estimate the period of the photometric variability using the Lomb-Scargle method. We combined sectors five and six into a single data set as they represent an almost-continuous time series. We find a period of $1.007$\,d for sectors five and six, and $1.010$\,d for sector 32. When fitting all sectors simultaneously we find a period of $1.008$\,d, with the best fit sinusoidal model remaining in phase between 2018 and 2020. The light curves do not exhibit any significant signature of an eclipsing binary. The variations appear to be symmetric, distinct from short-period contact binaries that have sharper minima. Based on these data, and the radial velocities presented in the previous subsection, we can confidently rule out the presence of a short-period near-equal mass companion to AF Lep.

\subsection{Spectral energy distribution}
The near-infrared spectral energy distribution (SED) of AF Lep was required to convert the measured contrasts from the observations presented in Sect.~\ref{sec:obs} into fluxes. We used the method described in \citet{Nielsen:2019cb} and \citet{Duchene:2023dv} to fit a joint atmospheric-evolutionary model to the optical ({\it Tycho} and {\it Gaia}; \citealp{Hog:2000wk,GaiaCollaboration:2021ev}) and near-infrared (Two Micron All Sky Survey, 2MASS; \citealp{Skrutskie:2006hla}) photometry for the star. We used a prior on the age of the star of $24\pm3$\,Myr \citep{Bell:2015gw}, and fixed the extinction to zero (Sect.~\ref{sec:alternatives}). We added in quadrature to the photometric uncertainties an error inflation term of 0.035\,mag such that the reduced $\chi^2$ was close to one. We ran an Markov-chain Monte Carlo (MCMC) analysis to sample the posterior distribution of each fitted parameter, as described in \citet{Nielsen:2019cb}. The best fit mass of $1.13_{-0.04}^{+0.10}$\,$M_{\odot}$ was consistent within uncertainties to the value of $1.29\pm0.20$\,$M_{\odot}$ reported in \citet{Stassun:2019bo}. The derived effective temperature of $6140\pm47$\,K is consistent with previous measurements from detailed spectroscopic analyses \citep{Casagrande:2011ji}.

We randomly selected 100 MCMC samples to generate a representative set of SEDs from which synthetic photometry and spectra, and their corresponding uncertainties, could be estimated. Each SED was convolved with the filter response curves for each instrument (see Sect.~\ref{sec:IFS}) to generate synthetic photometry, and was also degraded to the resolution of the IFS observations to generate synthetic spectra. The median of each of these was used as the synthetic measurement, and the standard deviation used as the uncertainty. We derived magnitudes of $Y_{\rm IFS}=5.50\pm0.02$, $J_{\rm IFS}=5.29\pm0.02$, $H2=5.02\pm0.03$, $K1=5.00\pm0.03$, and $K2=5.00\pm0.03$ for AF Lep. Fluxes in these filters and for the IFS channels are given in Table~\ref{tbl:sed}.

\section{Observations and data reduction}
\label{sec:obs}
\begin{table*}
\caption{Observing log.}
\label{tbl:log}
 \centering
\begin{tabular}{cccccccc}
\hline\hline
UT (mid) & MJD & Target & Instrument Setup & $t_{\rm DIT}$ (s) & $n_{\rm DIT}$ & $n_{\rm EXP}$ & $t_{\rm total}$ (s)\\
\hline
2022-10-20T05:10:21 & 59872.22 & AF Lep & IRDIS/DB-K12 & 16 & 7 & 24 & 2688\\
& & & IFS/YJH & 32 & 14 & 6 & 2688\\
& & HD 35591 & IRDIS/DB-K12 & 16 & 4 & 16 & 1024\\
& & & IFS/YJH & 32 & 8 & 4 & 1024\\
\hline
\end{tabular}
\end{table*}

AF Lep and the calibrator star \object{HD 35591} were observed with the Very Large Telescope/SPHERE (Spectro-Polarimetric High-contrast Exoplanet REsearch; \citealp{Beuzit:2019dl}) on the night of 2022 October 20. The star was observed simultaneously with the Infra-Red Dual-bean Imaging and Spectroscopy (IRDIS; \citealp{Dohlen:2008eu}), and the Integral Field Spectrograph (IFS; \citealp{Claudi:2008dj}) subcomponents. The observations were taken in the new star-hopping mode \citep{Wahhaj:2021kf}, where observations of the science targets are interleaved with rapid offsets to a calibrator star to obtain a reference image of the point spread function (PSF). An observing log is given in Table~\ref{tbl:log}. The observing conditions were excellent, with a typical Differential Imaging Motion Monitor (DIMM) seeing of $0\farcs46$ and coherence time of 6.7\,ms. The observations consisted of an off-axis measurement of AF Lep with a neutral density filter for flux calibration, an observation of both AF Lep and HD 35591 with satellite spots created by applying a set of sine waves to the deformable mirror to determine the position of the star behind the coronagraph, with the rest of the time devoted to the exoplanet search observations. The full observing sequence consisted of five sets of data on AF Lep, and four interleaved sets on the PSF calibrator star HD~35591. The observations were taken at a low elevation while the target was still rising; as such, the parallactic angle changed by only $\sim$4.5\,deg over the full sequence. Several exposures were taken with the adaptive optics loop opened. These were identified automatically via cross-correlation and were discarded.

The data for both stars were reduced using the SPHERE pipeline (v0.42.0), in conjunction with a set of tools and utilities developed to refine the output of the pipeline \citep{Vigan:2020vi}\footnote{\url{https://github.com/avigan/SPHERE}}. Together, these reduced the data by subtracting the dark current, correcting for the flat field, fixing bad pixels, and correcting for detector anamorphism. For the IFS data, the pipeline measures the position of the micro-spectra and calculates a wavelength solution using observations of an internal laser source. The off-axis observations of AF Lep were calibrated using the detector integration time and the throughput of the neutral density filter. The star position in each of the center frames was measured by fitting a Gaussian to the four satellite spots to determine the intersection of the lines connecting the two diagonal pairs. The star was assumed to be stationary behind the coronagraph during the full sequence, and all of the observations were aligned to this common center.

\subsection{SPHERE/IRDIS}
\label{sec:IRDIS}
\begin{figure*}
    \centering
    \includegraphics[width=0.8\textwidth]{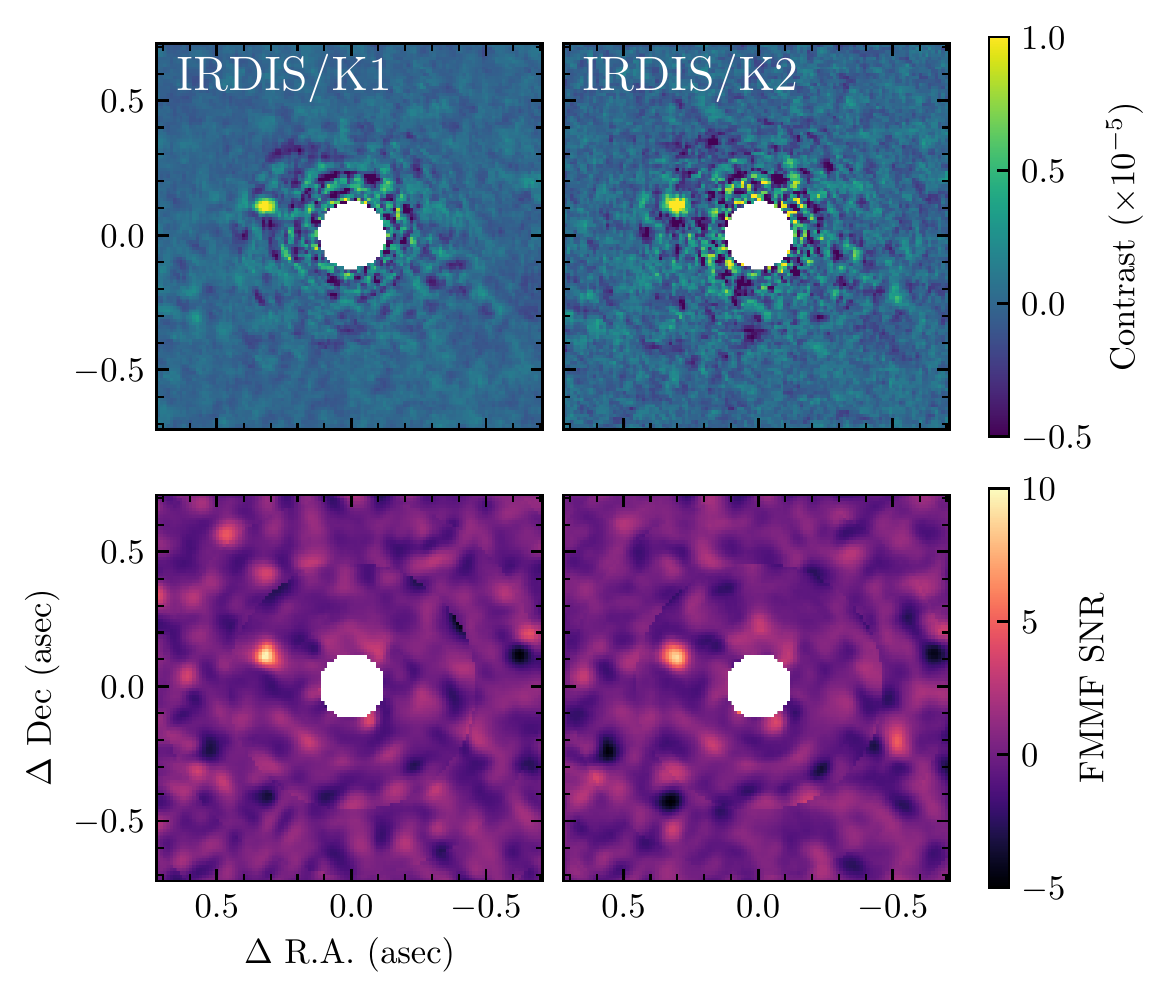}
    \caption{\label{fig:detection}Detection and significance of the detection of AF Lep b with IRDIS. (top row) Residual maps after KLIP processing for the K1 (left) and K2 (right) IRDIS data sets. The inner region within 10\,px is digitally masked. The companion is clearly visible at a position angle of 70 degrees. (bottom row) FMMF S/N maps for the K1 (left) and K2 (right) data sets. The companion is detected at an S/N of 10.5 and 9.4 in the two images.}
\end{figure*}
We used \texttt{pyKLIP} \citep{Wang:2015th}, an implementation of the Karhunen--Lo\'{e}ve image processing algorithm (KLIP; \citealp{Soummer:2012ig, Pueyo:2016hl}), to model and subtract the residual PSF in the IRDIS images not suppressed by the coronagraph. The observations in the two filters were treated separately. Each image was divided into seven annuli logarithmically spaced between an inner working angle of 10\,px and the edge of the cropped image ($218\times218$\,px), each divided into four segments. After the initial reduction, these segments were rotated to place the companion at the center of a segment. For each image of AF Lep, a reference PSF was constructed from a library containing all of the observations of the PSF calibrator HD 35591 in the same filter, and those observations of AF Lep in the same filter for which an astrophysical source would have rotated more than two pixels. The PSF was constructed from this reference library using 25 Karhunen--Lo\'{e}ve (KL) modes. The resulting PSF-subtracted images were then temporally averaged to produce the final images shown in Fig.~\ref{fig:detection} (top row) showing the clear detection of the companion.

Detection maps were calculated using a forward-model matched-filter (FMMF) approach, as outlined in \citet{Ruffio:2017ev}. This involved a pixel-wise cross-correlation with a forward-modeled PSF to account for the distorting effect of the KLIP processing on any point source within the image. The resulting S/N maps are shown in Fig.~\ref{fig:detection} (bottom panel), clearly showing the detection of the companion. We find an S/N of 10.5 in K1 and 9.4 in K2, these correspond to a 7.0-$\sigma$ and 6.5-$\sigma$ detection given the angular separation of the companion \citep{Mawet:2014ga}. The FMMF contrast curves for the IRDIS data sets are discussed in Sect.~\ref{fmmf-con}. We also used \texttt{PlanetEvidence} \citep{Golomb:2021ge} to assess the significance of the detection. We found an evidence ratio of $\ln\left(B_{10}\right)=4.6$ in favor of the planet model for the K1 detection, and $\ln\left(B_{10}\right)=3.5$ for the K2 detection. The thresholds for what constitutes a strong detection vary between authors, but the evidence ratio for the planet model in the K1 image would be seen as strong by most.

\subsection{SPHERE/IFS}
\label{sec:IFS}
\begin{figure*}
    \centering
    \includegraphics[width=1.0\textwidth]{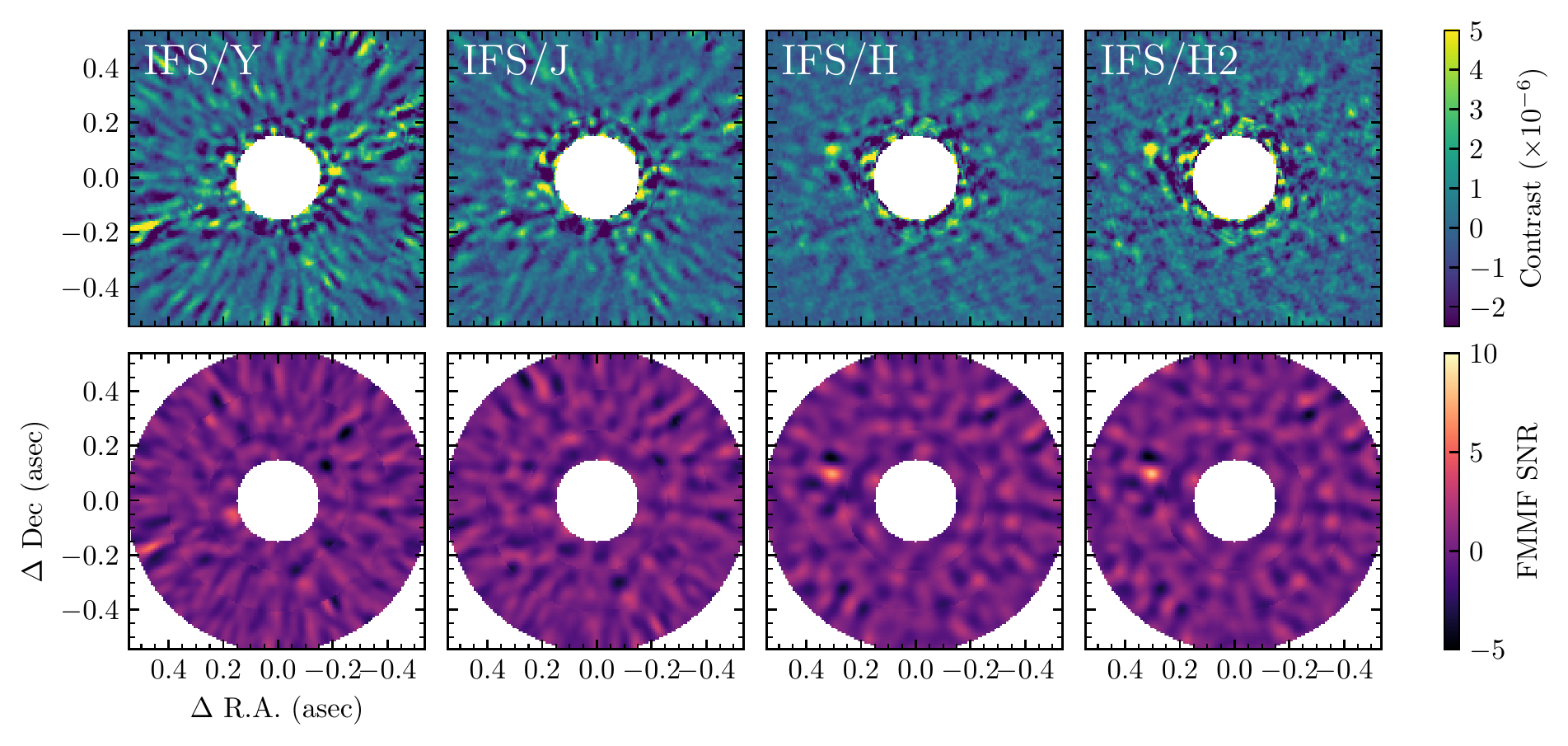}
    \caption{\label{fig:detection-ifs}As Fig.~\ref{fig:detection}, but for the four synthesized IFS data sets. Note that the contrast (top row) color scale is reduced compared to Fig.~\ref{fig:detection}, whereas the FMMF S/N is the same.}
\end{figure*}
\begin{figure}
    \centering
    \includegraphics[width=0.5\textwidth]{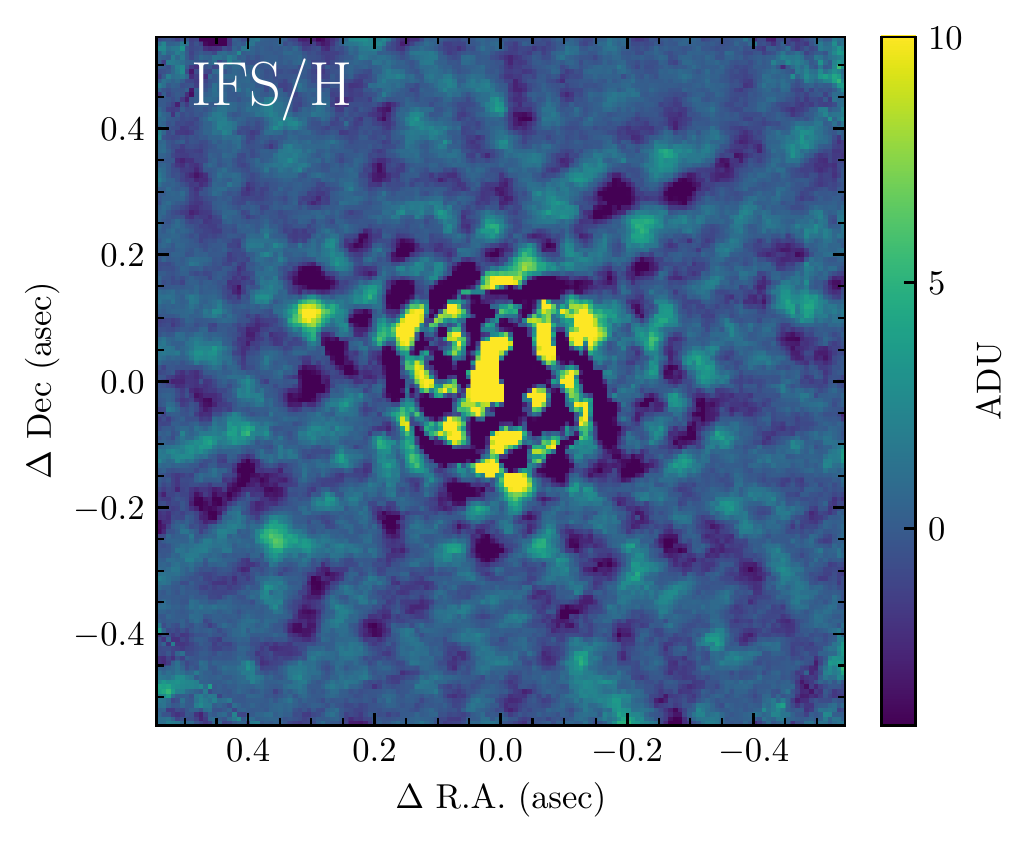}
    \caption{\label{fig:ifs-zwa}Final PSF-subtracted image of the $H$-band portion of the IFS data set, reduced using the pipeline presented in \citet{Wahhaj:2021kf}.}
\end{figure}
The IFS data were processed in a similar fashion to the IRDIS data as described in Sect.~\ref{sec:IRDIS}. We used five logarithmically spaced annuli spaced between 15\,px and 100\,px. Unlike for the IRDIS data, the IFS data did not have to be cropped. We processed the $Y$, $J$, and $H$ bands separately using only the channels within the cut-on and cut-off wavelengths of the standard $YJH$ filters ($0.966$--$1.121$\,$\mu$m for $Y$, $1.163$--$1.345$\,$\mu$m for $J$, and $1.474$--$1.651$\,$\mu$m for $H$). We also performed a reduction using just the three channels corresponding to the IRDIS $H2$ filter ($1.579$--$1.614$\,$\mu$m) to allow for a comparison of the photometry with other substellar companions studied with SPHERE. A model PSF was constructed for each input image from the reference library of images of both AF Lep and HD 35591 using ten KL modes. We did not use any of the spectral diversity when creating the reference library due to the presence of a strong striping pattern seen in the reduced data cubes that left a strong residual in the PSF-subtracted images when spectral differential imaging was used (e.g., \citealp{Berdeu:2020gd, Wahhaj:2021kf}). The resulting PSF-subtracted images were averaged in both time and wavelength to produce the final residual images shown in Fig.~\ref{fig:detection-ifs} (top row).

Detection maps  were calculated in the same way as for IRDIS and are shown in Fig.~\ref{fig:detection-ifs}. The companion is only detected within the $H$ band, although a low-significance (S/N$\sim$2.5) source can be seen at the same position within the $J$-band image. We estimate the S/N of the detections to be 7.3 at $H_{\rm IFS}$ and 8.4 at $H2$, corresponding to 5.9-$\sigma$ and 6.5-$\sigma$ at that angular separation \citep{Mawet:2014ga}. Due to the lower significance of the detection, we also reduced the IFS data using the pipeline described in \citet{Wahhaj:2021kf} that was developed especially for SPHERE reference-star differential imaging data sets. Using this pipeline we are able to detect AF Lep b in the $H$ band at an S/N $\sim$6 (see Fig.~\ref{fig:ifs-zwa}), with consistent astrometry and photometry to the measurements reported in Table~\ref{tbl:astro_photo}. We were not able to recover the companion at a significant level at either $Y$ or $J$ using this pipeline.

\subsection{Contrast curves}
\label{fmmf-con}
\begin{figure}
    \centering
    \includegraphics[width=0.5\textwidth]{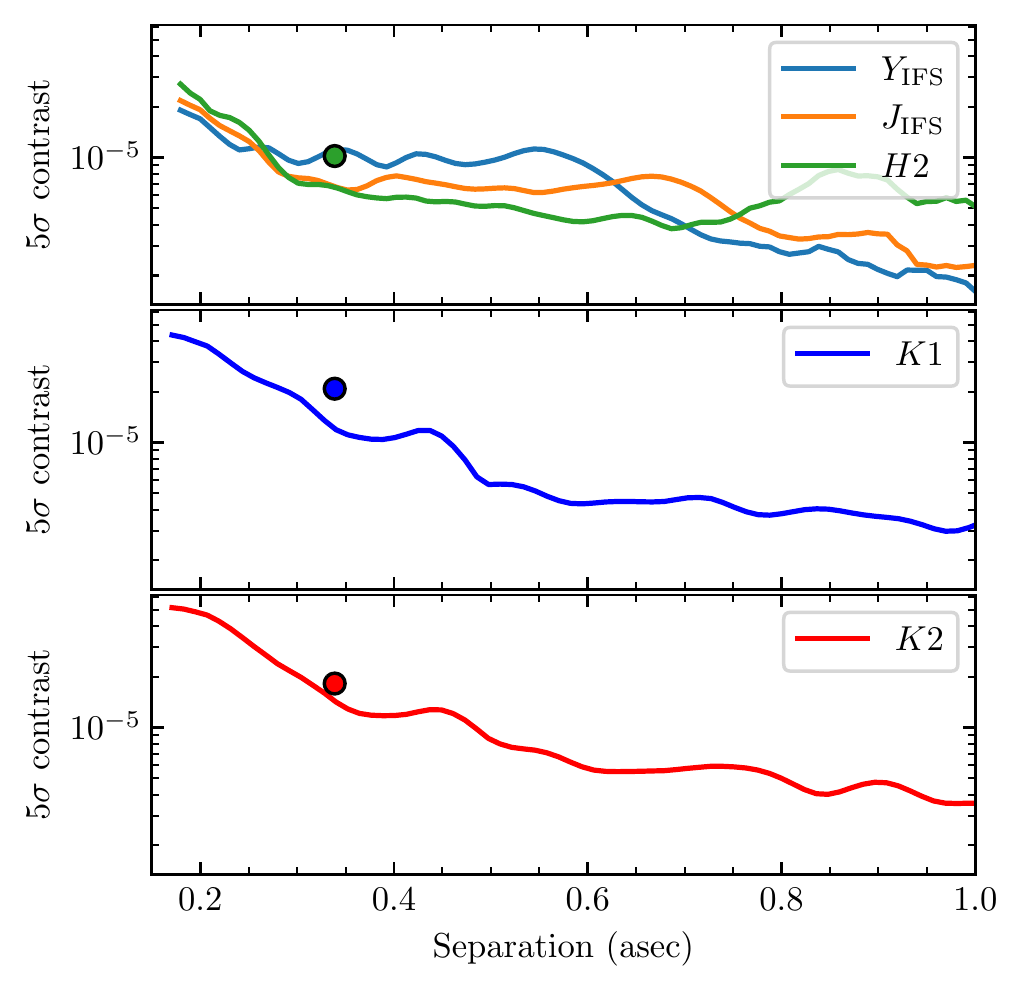}
    \caption{\label{fig:ccurve-irdifs}Azimuthally averaged sensitivity of the IFS (top), IRDIS K1 (middle), and K2 (bottom) data sets to companions as a function of angular separation. The measured contrast of AF Lep b within each data set is plotted (circle symbols).}
\end{figure}
The sensitivity of the IRDIS and IFS observations were calculated based on the approach outlined in \citep{Ruffio:2017ev}. The FMMF sensitivity maps described previously were calibrated using injection and recovery of sources at known contrasts. For the IFS data, we used as a template spectrum the same cloudy 1300\,K atmosphere model used in \citet{Ruffio:2017ev}. Sources were injected into the reduced IRDIS frames at four position angles (25, 115, 205, and 295\,deg), and nine separations between 15 and 95\,px at 10\,px increments. For IFS we used a finer sampling of six position angles between 0 and 75, alternating between ten injected sources between 25 and 180\,px and nine between 34 and 178\,px, both at 18\,px increments. This interleaved pattern gave a finer sampling of the algorithm throughput as a function of separation without the injected sources interfering with one another. The contrast of the injected sources was chosen in an iterative process to result in an S/N of approximately 10 in the reduced image. The data with injected sources were processed using the same \texttt{pyKLIP} parameters described in Sects.~\ref{sec:IRDIS} and \ref{sec:IFS}. This process was repeated six times, each with the pattern of simulated planets clocked by fifteen degrees. The ratio of the injected to recovered contrast was measured by averaging over all position angles, and used to calibrate the S/N and contrast maps. The resulting one-dimensional calibrated contrast curves are shown in Fig.~\ref{fig:ccurve-irdifs}, and are consistent with the estimated S/N of the planet detections.

\section{AF Lep b}
\label{sec:results}
\begin{table*}
\caption{Astrometric and photometric results.}
\label{tbl:astro_photo}
 \centering
\begin{tabular}{cccccccc}
\hline\hline
Band & $\rho$ & $\theta$ & Flux ratio & Flux & $\Delta$ mag & App. mag & Abs. mag\\
& (mas) & (deg) & $(\times 10^{-6})$ & ($\times 10^{-17}$ W\,m$^{-2}$\,$\mu$m) & & & \\
\hline
$Y_{\rm IFS}$ & $\cdots$        & $\cdots$       & $<10.9$      & $<38.74$ & $>12.40$ & $>17.92$ & $>15.77$\\
$J_{\rm IFS}$ & $\cdots$        & $\cdots$       & $<6.9$       & $<15.76$ & $>12.91$ & $>18.18$ & $>16.04$\\
$H2$          & ($328.8\pm4.1$) & ($70.1\pm0.7$) & $10.2\pm1.2$ & $12.70\pm1.52$ & $12.48\pm0.13$ & $17.50\pm0.13$ & $15.35\pm0.13$\\
\hline
$K1$          & $339.4\pm6.3$   & $70.3\pm1.1$   & $20.9\pm1.6$ & $9.78\pm0.79$ & $11.70\pm0.08$ & $16.70\pm0.09$ & $14.56\pm0.09$\\
$K2$          & $338.0\pm6.5$   & $70.3\pm1.1$   & $18.3\pm1.9$ & $6.68\pm0.71$ & $11.84\pm0.11$ & $16.85\pm0.12$ & $14.70\pm0.12$\\
\hline
Adopted & $338.7\pm6.4$ & $70.3\pm1.1$ & $\cdots$ & $\cdots$ & $\cdots$ & $\cdots$\\
\hline
\end{tabular}
\end{table*}

\subsection{Astrometry}
\label{sec:astrometry}
The relative astrometry between the host star and the companion was measured using Bayesian KLIP-FM astrometry (BKA; \citealp{Wang:2016gl}). This algorithm forward-models the effects of the KLIP processing on the PSF in the image, providing for a better match between the companion and the model PSF used to fit its location. We used broadly the same KLIP parameters as for the PSF reduction described previously, except the subtraction was only performed within a 20-pixel radius 90-degree annulus segment, centered on the location of the companion. The forward modeled PSF was fit to the companion using an MCMC approach to sample the posterior distributions for the companion position and relative brightness. Fit residuals are shown in Fig.~\ref{fig:bka}. Pixel offsets were converted to sky-plane separations and position angles using the calibration values given in \citet{Maire:2021bg} for IRDIS, and the SPHERE User Manual\footnote{\url{https://www.eso.org/sci/facilities/paranal/instruments/sphere/doc.html}} for IFS. We combined our fitting uncertainties and calibration uncertainties in quadrature with a 0.5\,px uncertainty on the location of the star. The photometry and relative astrometry from each image, and a weighted average of the astrometry that we adopt for subsequent analyses, are given in Table~\ref{tbl:astro_photo}. We do not include the IFS measurement in the average due to the lower S/N of the detection and the significant structure still visible within the residuals (see Appendix~\ref{ap:bka}).

\begin{figure}
    \centering
    \includegraphics[width=0.5\textwidth]{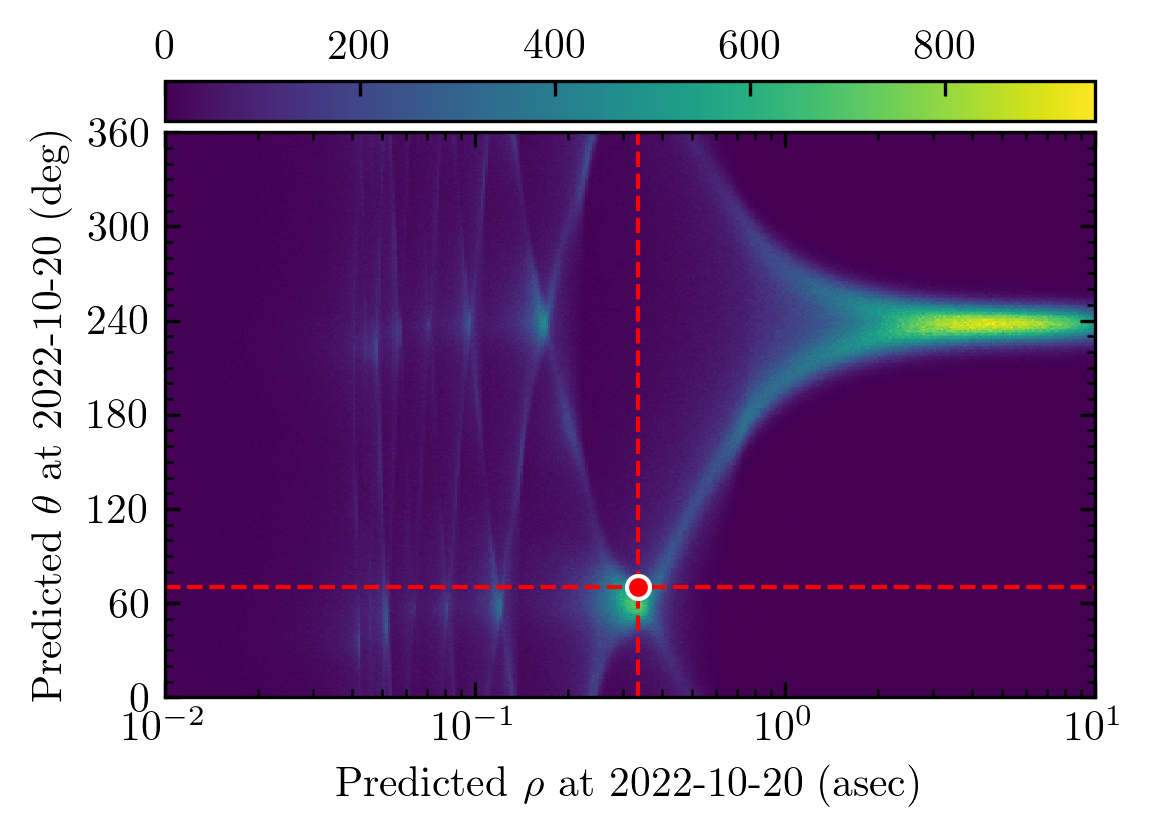}
    \caption{\label{fig:prediction}Two-dimensional histogram of the predicted separation and position angle at the time of the SPHERE observations of the companion orbits selected by the rejection sampling analysis that are consistent with the measured astrometric acceleration. The measured position of AF Lep b is at the intersection of the two dashed lines.}
\end{figure}
Figure~\ref{fig:prediction} shows a comparison of the measured separation and position angle of AF Lep b is compared to predicted position of simulated companions that were found using the rejection sampling analysis described previously. The companion's separation and position is nearly coincident with a large population of samples, differing by about ten degrees from the mode of that population. This difference may not be significant; it is likely a result of the rejection sampling analysis only considering circular orbits rather than indicative of a miscalibration of the proper motion catalog used in this analysis. AF Lep b is almost exactly where it was predicted to be from the astrometric signal. The large population of samples at a position angle of $\sim$240\,deg at separations beyond an arsecond are either high mass brown dwarfs or stellar mass (Fig.~\ref{fig:p-m}). These were already confidently excluded based on previous imaging searches, and can again be excluded using the IRDIS observations presented in this work.

We used the software package \texttt{orvara} \citep{Brandt:2021ek} to perform a simultaneous fit of the relative astrometry presented here, the proper motion differences reported in the HGCA, and the radial velocity measurements from \citet{TalOr:2019gm}. We sampled the posterior distributions of the orbital elements using 128 walkers advanced for 500,000 steps at each of five temperatures. The orbital semimajor axis ($a=9.3_{-1.8}^{+2.4}$\,au) and companion mass ($M_2 = 4.3_{-1.2}^{+2.9}$\,$M_{\rm Jup}$) were well constrained even with the single epoch of relative astrometry, corresponding to an orbital period of $24.8_{-6.8}^{+10.0}$\,yr. The orbital eccentricity ($e=0.32_{-0.23}^{+0.41}$) and inclination ($i=98_{-42}^{+32}$\,deg) were only marginally constrained. The posterior distributions of these parameters are shown in Fig.~\ref{fig:orvara-pdf}.

\subsection{Near-infrared photometry and colors}
\begin{table}
\caption{Near-infrared colors of AF Lep b.}
\label{tbl:colors}
 \centering
\begin{tabular}{cc}
\hline\hline
& Color\\
\hline
$J_{\rm IFS}-K1$ & $>1.48$ \\
$J_{\rm IFS}-K2$ & $>1.34$ \\
$H2-K1$ & $0.80\pm0.16$ \\
$H2-K2$ & $0.65\pm0.18$ \\
$K1-K2$ & $-0.15\pm0.15$\\
\hline
\end{tabular}
\end{table}

\label{sec:photo}
\begin{figure*}
    \centering
    \includegraphics[width=\textwidth]{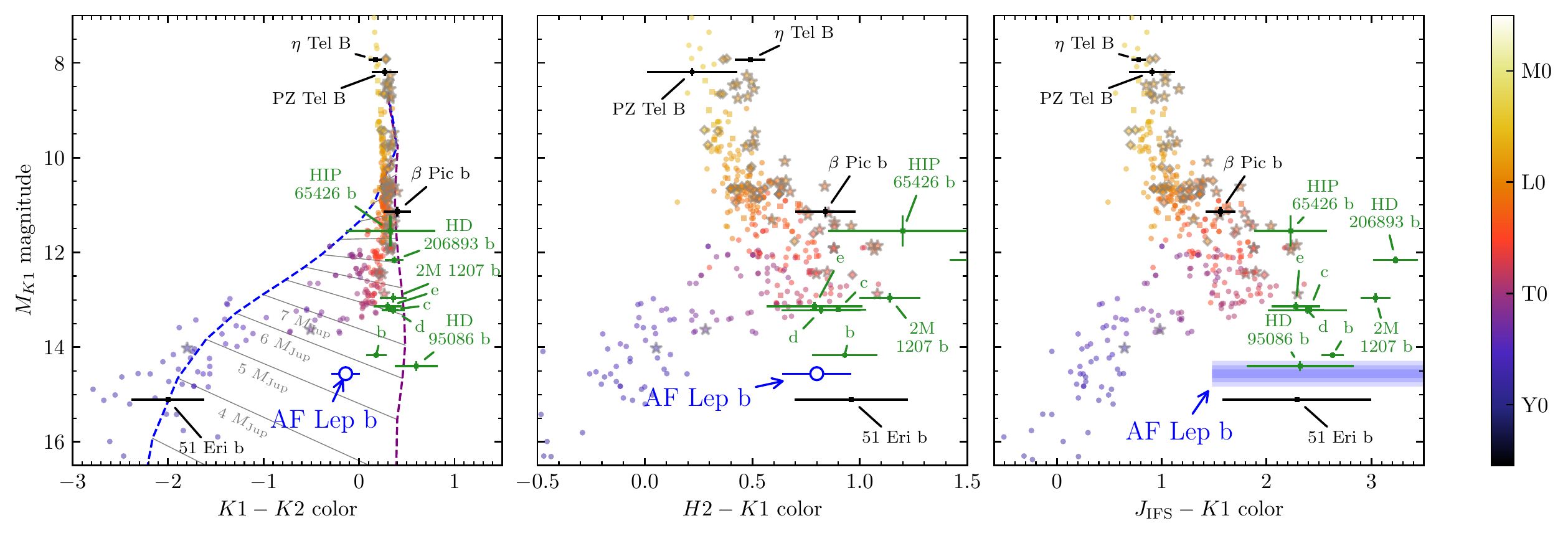}
    \caption{\label{fig:cmd}Near-infrared color-magnitude diagrams showing AF Lep b (blue) and other young imaged companions (other $\beta$~Pic moving group members in black, others in green) compared to the field-gravity (circles), low-gravity (diamonds), and very low-gravity (stars) brown dwarfs, colored according to their spectral type. The HR 8799 planets are indicated by their letter. Evolutionary models interpolated to 24\,Myr from the \texttt{AMES-Cond} (blue; \citealp{Chabrier:2000hq}) and \texttt{AMES-Dusty} (purple; \citealp{Baraffe:2003bj}) grids,  from which masses were derived, are plotted in the left panel. Companion photometry from \citet{Patience:2010hf}, \citet{Bonnefoy:2014dh}, \citet{Maire:2016go}, \citet{Zurlo:2016hl}, \citet{Chauvin:2017ev}, \citet{Chilcote:2017fv}, \citet{Samland:2017jh}, \citet{Chauvin:2018ib}, \citet{Langlois:2021ff}, \citet{WardDuong:2021ev}, and  \citet{BrownSevilla:2022ww}.}
\end{figure*}

The near-infrared photometry of AF Lep b was obtained from forward modeling analysis described in Sect.~\ref{sec:astrometry}. These flux ratios, the derived magnitude differences, and the apparent and absolute magnitudes are given in Table~\ref{tbl:astro_photo}. For the $Y$- and $J$-band limits, we use the 5$\sigma$ contrasts at the separation of AF Lep b as described in Sect.~\ref{fmmf-con}. Color measurements and lower limits are given for various combinations of bandpasses in Table~\ref{tbl:colors}.

The photometry and colors of AF Lep b are compared to those of isolated brown dwarfs of a range of surface gravities, and of young, imaged planetary-mass companions in Fig.~\ref{fig:cmd}. Comparison object photometry was computed from near-infrared spectra obtained from the SpeX Prism Library \citep{Burgasser:2014tr}, the IRTF Spectral Library \citep{Cushing:2005ed}, and the Montreal Spectral Library (e.g., \citealp{Gagne:2015dc, Robert:2016gh}). The spectra were normalized to literature near-infrared photometry and convolved with the response of the IRDIS filters ($K1$ and $K2)$ and the synthetic IFS filters ($Y_{\rm IFS}$, $J_{\rm IFS}$, $H2$) to create synthetic photometry. Parallax measurements for the substellar objects were from \citet{Dupuy:2012bp}, \citet{Dupuy:2013ks}, and \citet{Liu:2016co} (and references therein). References for the highlighted companion photometry is given in the caption of Fig.~\ref{fig:cmd}.

Considering only the $K$-band photometry plotted in Fig.~\ref{fig:cmd} (left panel), we find that AF Lep b appears displaced blue-ward from the sequence of imaged companions that extends from the inflection point seen near the L--T spectral type transition down to the location of \object{HR 8799 b} and \object{HD 95086 b}. These objects have been noted for their significant under-luminosity with respect to other L--T transition objects (e.g., \citealp{Chauvin:2017ev}), which can be explained by invoking thick photospheric clouds and disequilibrium chemistry (e.g., \citealp{Currie:2014fm} for HR 8799 b). The significantly bluer $K1-K2$ color of AF Lep could be explained by a lower effective temperature leading to enhanced methane absorption depressing the flux beyond $2.2$\,$\mu$m, although variations in near-infrared color can also be induced by changes in the vertical distribution of clouds without changing the effective temperature (e.g., \citealp{Barman:2011fe}).

We also used the $H$-band detection and $J$-band upper limit to compare the $H2-K1$ and $J_{\rm IFS}-K$ colors of AF Lep. The $H-K$ color is consistent with HR 8799 b, albeit with a significantly lower $K$-band luminosity (Fig.~\ref{fig:cmd}, middle panel). It also shares a similar $H2-K1$ color to 51 Eri b, which is significantly redder than other T dwarfs of a similar spectral type. This is due to the sensitivity of this color index to surface gravity at these effective temperatures \citep{Samland:2017jh}; at lower surface gravities the effect of collision-induced absorption is reduced, leading to a much redder $H-K$ color (e.g., \citealp{Knapp:2004ji}). The $J_{\rm IFS}-K1$ lower limit is consistent with both HR 8799 b and HD 95086 b (Fig.~\ref{fig:cmd}, right panel), and is inconsistent with objects earlier than $\sim$L0 (see Sect.~\ref{sec:alternatives}).

\begin{table}
\caption{Model-derived parameters of AF Lep b.}
\label{tbl:model_parameters}
 \centering
\begin{tabular}{cccc}
\hline\hline
Parameter & Unit & {\texttt AMES-Dusty} & {\texttt AMES-Cond}\\
\hline
$M$ & $M_{\rm Jup}$ & $6.1\pm0.3$ & $4.1\pm0.3$\\
$R$ & $R_{\rm Jup}$ & $1.38\pm0.01$ & $1.28\pm0.01$\\
$L$ & $L_{\odot}$ & $-4.75\pm0.02$ & $-5.16\pm0.02$\\
$T_{\rm eff}$ & $K$ & $1030\pm12$ & $820\pm10$ \\
$\log (g)$ & [dex] & $4.00\pm0.04$ & $3.80\pm0.04$\\
\hline
\end{tabular}
\end{table}

We estimated a mass of the companion by comparing the two $K$-band photometric measurements to evolutionary models interpolated to the estimated age of the host star. We used the \texttt{AMES-Cond} \citep{Chabrier:2000hq} and \texttt{AMES-Dusty} \citep{Baraffe:2003bj} model grids that have pre-computed fluxes in the SPHERE filters. A rejection sampling-based approach was used to transform the photometric measurements into a mass. We generated $10^7$ samples uniformly in age (between 1 and 50\,Myr) and mass (between 1 and 15\,$M_{\rm Jup}$) for which synthetic $K1$ and $K2$ photometry were computed. A $\chi^2$ was calculated for each sample based on a comparison of the synthetic photometry to the two measurements, which was used to perform the rejection sampling. This was repeated for the two model grids, yielding a mass of $4.1\pm0.3$\,$M_{\rm Jup}$ for the \texttt{AMES-Cond} grid, and $6.1\pm0.3$\,$M_{\rm Jup}$ for the \texttt{AMES-Dusty} grid. These are both consistent with the dynamical mass measurement presented in Sect.~\ref{sec:astrometry}. The other model-derived parameters of AF Lep b are reported in Table~\ref{tbl:model_parameters}, the uncertainties on each only include the statistical uncertainty on the absolute near-infrared photometry and on the age of the system.

\subsection{Spectroscopy}
\label{sec:spectroscopy}
\begin{table}
\caption{Spectral energy distribution of AF Lep and AF Lep b.}
\label{tbl:sed}
\centering
\begin{tabular}{cccc}
\hline\hline
$\lambda$ & Contrast & $F_{\lambda} (A)$ & $F_{\lambda} (b)$ \\ 
$\mu$m & $\times 10^{-6}$ & $\times 10^{-12}$ & $\times 10^{-17}$ \\
 & & W\,m$^{-2}$\,$\mu$m & W\,m$^{-2}$\,$\mu$m\\
\hline
$0.966$ & $<9.27$ & $41.87\pm0.64$ & $<38.82$ \\
$0.979$ & $<10.14$ & $40.76\pm0.63$ & $<41.4$ \\
$0.994$ & $<9.72$ & $39.35\pm0.61$ & $<38.22$ \\
$1.011$ & $<12.30$ & $37.86\pm0.60$ & $<46.62$ \\
$1.030$ & $<11.97$ & $36.36\pm0.58$ & $<43.56$ \\
$1.048$ & $<9.69$ & $34.86\pm0.56$ & $<33.75$ \\
$1.067$ & $<8.37$ & $33.25\pm0.54$ & $<27.81$ \\
$1.085$ & $<6.54$ & $31.79\pm0.53$ & $<20.85$ \\
$1.102$ & $<6.27$ & $30.71\pm0.51$ & $<19.29$ \\
$1.121$ & $<4.20$ & $29.64\pm0.50$ & $<12.42$ \\
$1.143$ & $<4.83$ & $28.38\pm0.48$ & $<13.68$ \\
$1.163$ & $<4.32$ & $27.25\pm0.47$ & $<11.79$ \\
$1.181$ & $<5.79$ & $26.21\pm0.46$ & $<15.21$ \\
$1.200$ & $<7.62$ & $25.21\pm0.45$ & $<19.20$ \\
$1.220$ & $<6.45$ & $24.34\pm0.44$ & $<15.72$ \\
$1.239$ & $4.18\pm2.22$ & $23.50\pm0.43$ & $9.82\pm5.21$ \\
$1.259$ & $5.98\pm2.13$ & $22.56\pm0.42$ & $13.50\pm4.81$ \\
$1.279$ & $7.04\pm1.89$ & $21.57\pm0.41$ & $15.18\pm4.08$ \\
$1.298$ & $7.45\pm1.75$ & $20.88\pm0.40$ & $15.56\pm3.66$ \\
$1.315$ & $7.43\pm2.14$ & $20.34\pm0.39$ & $15.11\pm4.37$ \\
$1.329$ & $6.54\pm2.06$ & $19.83\pm0.38$ & $12.96\pm4.10$ \\
$1.345$ & $6.04\pm2.13$ & $19.30\pm0.38$ & $11.66\pm4.13$ \\
$1.371$ & $3.58\pm3.17$ & $18.39\pm0.37$ & $6.58\pm5.84$ \\
$1.404$ & $3.02\pm2.84$ & $17.31\pm0.36$ & $5.23\pm4.91$ \\
$1.424$ & $2.20\pm2.27$ & $16.68\pm0.35$ & $3.67\pm3.79$ \\
$1.440$ & $2.16\pm1.97$ & $16.18\pm0.34$ & $3.50\pm3.19$ \\
$1.457$ & $2.44\pm1.76$ & $15.69\pm0.34$ & $3.82\pm2.76$ \\
$1.474$ & $2.20\pm1.73$ & $15.22\pm0.33$ & $3.35\pm2.64$ \\
$1.492$ & $2.04\pm1.82$ & $14.76\pm0.32$ & $3.02\pm2.69$ \\
$1.509$ & $4.12\pm1.56$ & $14.34\pm0.32$ & $5.90\pm2.24$ \\
$1.527$ & $4.70\pm1.41$ & $13.96\pm0.31$ & $6.55\pm1.98$ \\
$1.544$ & $6.34\pm1.74$ & $13.55\pm0.31$ & $8.59\pm2.37$ \\
$1.561$ & $8.75\pm1.94$ & $13.10\pm0.30$ & $11.47\pm2.56$ \\
$1.579$ & $9.16\pm1.83$ & $12.63\pm0.30$ & $11.57\pm2.32$ \\
$1.597$ & $11.21\pm1.43$ & $12.22\pm0.29$ & $13.70\pm1.78$ \\
$1.614$ & $12.98\pm1.56$ & $11.87\pm0.28$ & $15.41\pm1.89$ \\
$1.629$ & $13.28\pm2.06$ & $11.55\pm0.28$ & $15.34\pm2.41$ \\
$1.641$ & $14.28\pm1.87$ & $11.28\pm0.27$ & $16.11\pm2.14$ \\
$1.651$ & $16.32\pm2.39$ & $11.07\pm0.27$ & $18.07\pm2.68$ \\
\hline
$1.033$ ($Y_{\rm IFS}$) & $<10.9$ & $35.54\pm0.57$ & $<38.74$ \\
$1.242$ ($J_{\rm IFS}$) & $<6.9$ & $22.84\pm0.42$ & $<15.76$ \\
$1.586$ ($H2$) & $10.2\pm1.2$ & $12.45\pm0.29$ & $12.70\pm1.52$ \\
$2.105$ ($K1$) & $20.9\pm1.6$ & $4.68\pm0.12$ & $9.78\pm0.79$ \\
$2.253$ ($K2$) & $18.3\pm1.9$ & $3.65\pm0.09$ & $6.68\pm0.71$ \\
\hline
\end{tabular}
\end{table}
The spectrum of AF Lep b was measured from the IFS data set using \texttt{extractSpec} \citep{Greenbaum:2018hz}, a module of \texttt{pyKLIP} that uses a forward model-based approach for spectral extraction \citep{Pueyo:2016hl}. A KLIP reduction was performed on a segment centered at the location of AF Lep b with a size of $\pm20$\,px and $\pm45$\,deg, using the same KLIP parameters as described in Sect.~\ref{sec:IFS}. A forward model of the stellar PSF was calculated at this position at each wavelength. This was combined with the PSF-subtracted images produced by KLIP to measure the contrast spectrum of the planet by finding the least-squares solution to Eq.~3 of \citet{Greenbaum:2018hz}. Rather than evaluating this over the full segment, the inverse problem was solved using a 20-pixel-width stamp centered on the location of the planet.

To estimate the uncertainty on the spectrum we performed a series of injection and recovery tests. We injected a source into the reduced data cubes prior to PSF subtraction that had the same contrast spectrum as measured for AF Lep b and then retrieved it using the same procedure. For the wavelength channels with $\lambda<1.2$\,$\mu$m the measured contrast of AF Lep b was negative, likely due to the residual speckle field and the intrinsic faintness of the planet. The contrast of the injected companion at these wavelengths was set to zero. This process was repeated at nine position angles at the same separation as the planet. We adopted the standard deviation of the recovered contrasts in each wavelength channel as the uncertainty to encapsulate the full range of recovered spectra.

The measured contrast between star and planet in each wavelength channel is given in Table~\ref{tbl:sed}, along with the spectrum of the host star and of the companion. The companion is only detected at a high significance in the last few channels of the $H$ band. There is a marginal detection of the companion at $J$, consistent with our detection limit in Table~\ref{tbl:astro_photo}, and there is no detection at $Y$. Because of the lack of even a marginal detection in the $Y$ band (e.g., Fig.~\ref{fig:detection-ifs}, left column), we exclude these channels from the subsequent analysis. Figure~\ref{fig:spec-res} shows the KLIP-processed images, the forward model multiplied by the contrast spectrum of the planet, and the residuals for each spectral channel. We note that the overall quality of the reduction is poor except from a few channels within the $H$ band, as is evident from the lack of a significant detection in the collapsed $J$-band data set (Fig.~\ref{fig:detection-ifs}). We are likely affected by residual speckles (positive, and negative from the reference star) that can introduce significant correlated noise into the extracted spectrum. As such, we only make make limited inferences on the properties of the companion using these data.

\begin{figure}
    \centering
    \includegraphics[width=0.5\textwidth]{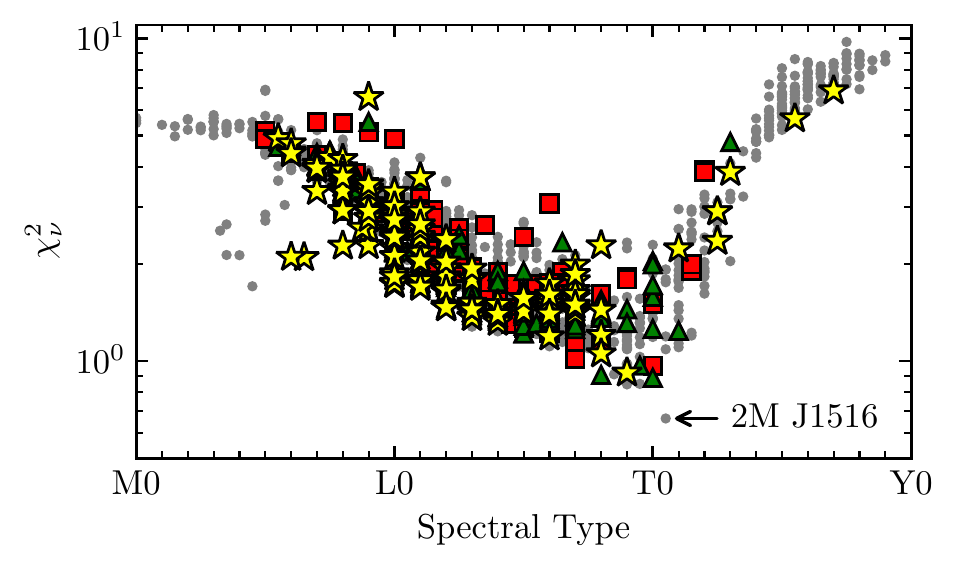}
    \caption{\label{fig:spec-chi2}Goodness of fit for the objects within the spectral library when compared to the photometry and spectroscopy of AF Lep b. Objects with gravity measurements are indicated: field gravity (red squares), intermediate (green triangles), and low gravity (yellow stars).}
\end{figure}
\begin{figure}
    \centering
    \includegraphics[width=0.5\textwidth]{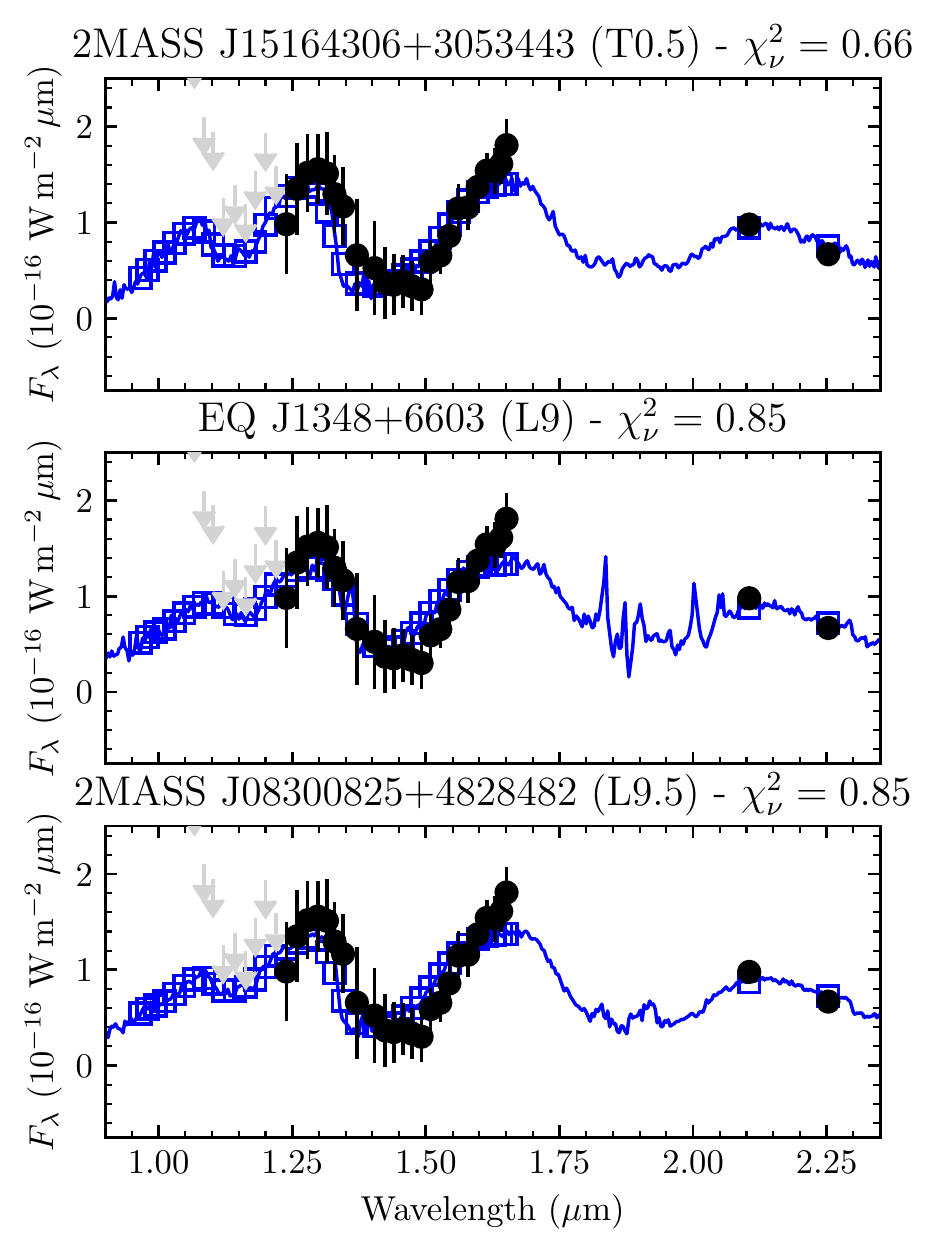}
    \caption{\label{fig:spec-compare}Three objects within the spectral library that best match the photometric and spectroscopic measurements of AF Lep b. The $Y$-band upper limits are not included in this analysis.}
\end{figure}

The photometry and spectroscopy of AF Lep b were compared to the library of substellar spectra described in Sect.~\ref{sec:photo}. For each object we calculated synthetic photometry in the IRDIS $K1$ and $K2$ filters, as well as synthetic spectra at the resolution of the IFS observations ($R\sim 30$). The optimal scaling factor that minimized the $\chi^2$ was calculated under the assumption that the uncertainty of the synthetic photometry and spectroscopy was negligible relative to that of AF Lep b. The goodness of fit for each object within the library is plotted as a function of spectral type in Fig.~\ref{fig:spec-chi2}. The best-fit objects are found near the L--T transition, consistent with the position of the companion on the color magnitude diagram (Fig.~\ref{fig:cmd}). The M dwarfs and mid to late T dwarfs within the library are a significantly worse fit.

A comparison between the spectra of the three best-fit objects and the measured SED of AF Lep b is shown in Fig.~\ref{fig:spec-compare}. The three objects are all at the L--T transition (Fig.~\ref{fig:spec-chi2}), and reproduce the large-scale features seen in the SED of AF Lep b; a significant absorption feature between $J$ and $H$, as well as the flux ratio between the $K1$ and $K2$ filters. The best-fitting object \object{2MASS J15164306+3053443} (2M 1516) was initially classified as a T0.5 dwarf by \citet{Chiu:2006jd} who noted that it appeared unusually red in the near-infrared compared to other objects of the same spectral type. It was found to have unusually red mid-infrared colors indicative of strong vertical mixing and thick photospheric clouds \citep{Leggett:2007if}. The object was found to be an irregular variable in the mid-infrared \citep{Metchev:2015dr}, suggestive of a rapid evolution of the visible photospheric features. The object is also noted for having unusually strong H$_2$O absorption bands \citep{Burgasser:2010df}, and weak CH$_4$ bands \citep{Chiu:2006jd}. Using a binary model with a blended spectra only offered a marginal improvement over a single-object template \citep{Chiu:2006jd,Burgasser:2010df}, so it is not yet known whether the spectral peculiarities are caused by either binarity or unusual atmospheric properties. 

\section{Alternative scenarios}
\label{sec:alternatives}
We have three independent lines of evidence to suggest that AF Lep b is a planetary-mass companion and not a background star. Firstly, the proper motion signal and photometry are both consistent with a $\sim$3--7 Mjup companion, and the position angle of the companion is consistent with the proper motion anomaly. Secondly, the spectrum and photometry are individually and jointly consistent with objects at the L--T transition, rather than a more distant star. Thirdly, the null-detection of background sources in previous high-contrast imaging searches exclude almost all plausible types of background objects. Nevertheless, we must entertain the possibility that this is indeed a chance alignment with an unassociated object such as a background star or a background or foreground brown dwarf.

AF Lep has galactic coordinates of ($l,b$) = ($214^{\circ}, -24^{\circ}$), toward the galactic anticenter and somewhat out of the galactic plane.  This is consistent with fewer background objects expected along the line of sight compared to a star close to the plane.  We retrieved a simulation of stars in a one square degree field in the direction of AF Lep, using the Besan\c{c}on simulation \citep{Robin:2003jk}, finding 1697 simulated stars brighter than $K=16.6$, corresponding to a number density of $1.3\times 10^{-4}$ stars per square arcsecond. Thus, a chance alignment with a background star at $\rho \sim 34$\,mas is unlikely with a probability of $\sim$$4.6\times10^{-5}$.  We do not expect any significant reddening of background objects due to the small amount of extinction in the direction of AF Lep \citep{Schlegel:1998fw}, with a total extinction of $E(g-r)=0.13$ (corresponding to $A_J\sim0.1$, $A_H\sim 0.06$, and $A_K\sim0.04$) out to 5\,kpc \citep{Green:2019go}.

\begin{figure}
    \centering
    \includegraphics[width=0.5\textwidth,trim={3.0cm 3.0cm 1.6cm 9cm}]{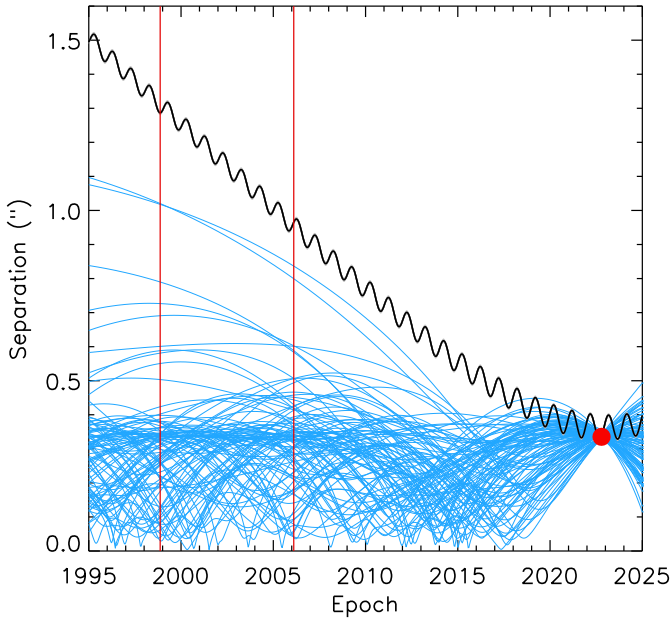}
    \caption{\label{fig:background}Expected motion from a stationary background object (black line and gray track), given the 2022 astrometry. We assume $\mu_\alpha=16.915$ mas/yr, $\mu_\delta=-49.318$ mas/yr and $\pi$=37.2539 mas.  Blue curves represent orbits drawn from the posterior distributions of the orbit fit. We expect a background object to have a significantly different trajectory in separation compared to an orbiting planet over the next year. The vertical red lines mark the epochs of HST/NICMOS and Keck/NIRC2 observations discussed in the text.}
\end{figure}
\begin{figure}
    \centering
    \includegraphics[width=0.35\textwidth]{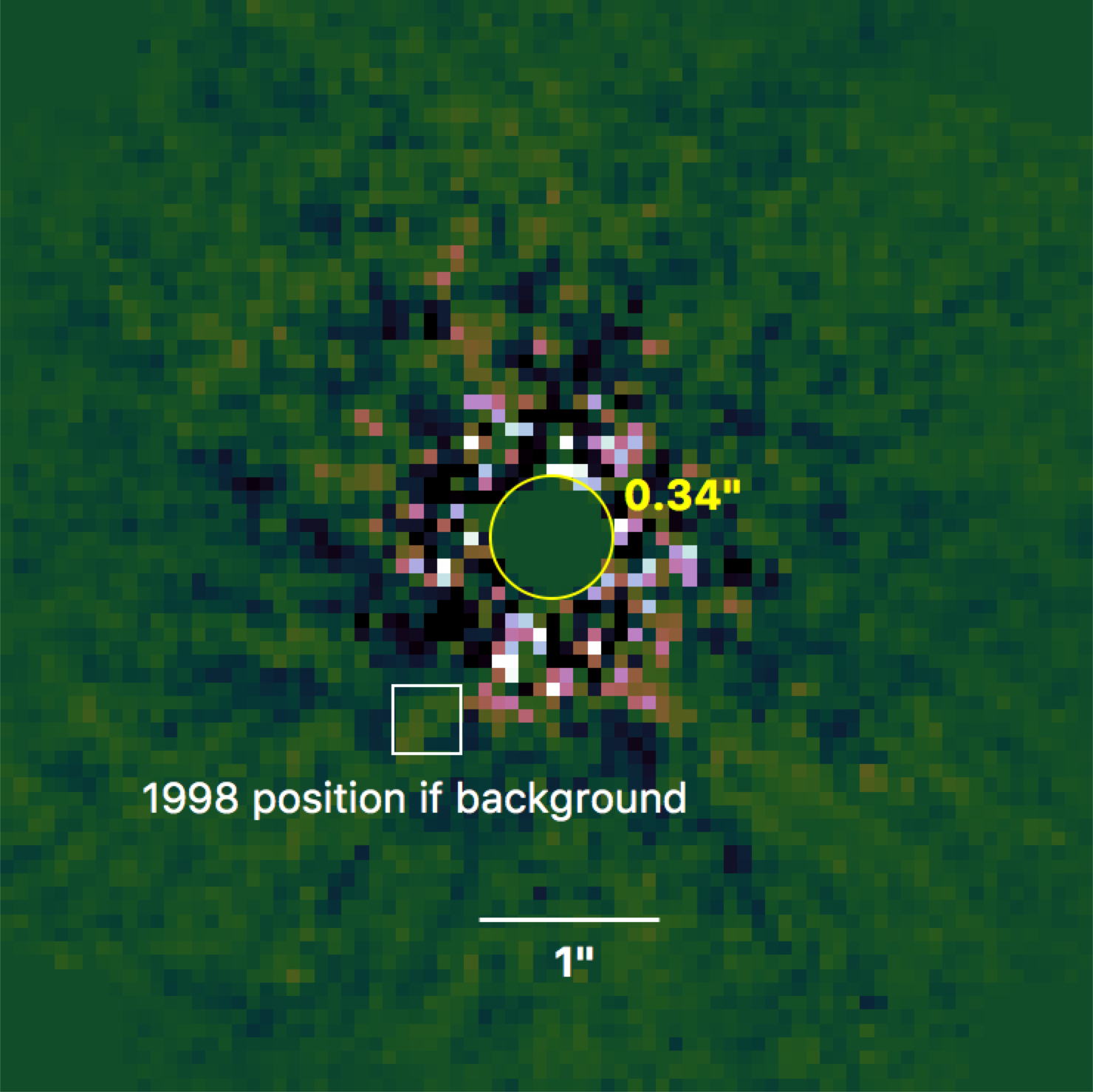}   \caption{\label{fig:hst_alice} F160W NICMOS field from 1998 as processed by the ALICE pipeline \citep{Choquet:2014ej,Hagan:2018ij}. Since the $\sim$25-year orbital period of AF Lep b is comparable to the 24-year difference between these HST observations and our new data, the planet probably lies at a similar position relative to the star in both data sets.  However, we do not expect a detection of the planet with HST NICMOS because the occulting spot has radius $0\farcs3$ and the contrast needed to achieve a detection begins past $0\farcs5$ radius from the star. The white box indicates the expected position of the companion in this 1998 epoch if it were a stationary background star instead of a common proper motion companion.}
\end{figure}
\begin{figure}
    \centering
    \includegraphics[width=0.5\textwidth]{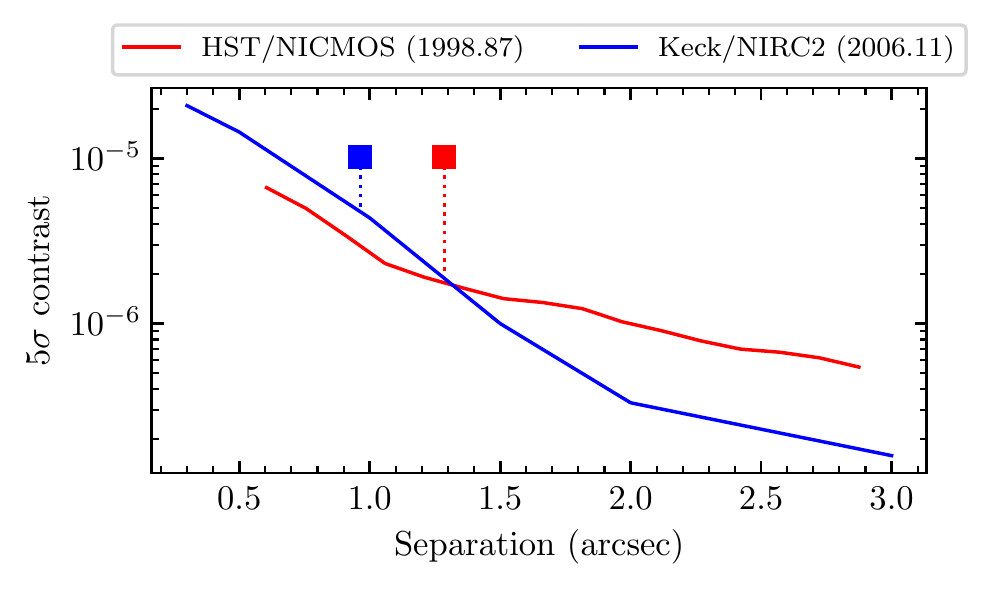}   \caption{\label{fig:literature-contrast}Contrast curves from two previous high-contrast imaging observations of AF Lep \citep{Galicher:2016hg,Hagan:2018ij}. The predicted separation if the companion was in fact a stationary background star is indicated for each epoch, significantly above the contrast curve in both cases.}
\end{figure}

Figure~\ref{fig:background} shows the expected motion of a stationary background object from the discovery epoch in 2022, given the \textit{Gaia} proper motion, parallax, and uncertainties for AF Lep. The null detections in previous high-contrast imaging searches from late 1998 (Fig.~\ref{fig:hst_alice}; \citealp{Hagan:2018ij}) and in early 2006 \citep{Galicher:2016hg} allow us to confidently reject a distant background star. The contrast curves and predicted location of a stationary background source for both epochs are shown in Fig.~\ref{fig:literature-contrast}: in both cases if the candidate were a distant background star, it would be well above the contrast curves. The {\it Hubble} Space Telescope (HST) observations made in late 1998 with Near Infrared Camera and Multi-Object Spectrometer (NICMOS; see Appendix~\ref{ap:hst}) achieved a contrast of $\Delta H=14.2$\,mag \citep{Hagan:2018ij} at the predicted separation of a stationary background star compared to the $\Delta H\sim12.5$\,mag measured for the companion in the IFS data set. The Keck observations made with the Near Infrared Camera 2 (NIRC2) in early 2006 achieved a contrast of $\Delta m=13.4$\,mag at the background position in the CH$_4$-short filter, which shares a similar effective wavelength but almost double the effective width as the IRDIS $H2$ filter. As with the NICMOS data set, a distant background star lies well above the published contrast curve given the measured contrast in the IFS data set. To remain undetectable in both of these data sets a background star would have to have a total proper motion of between 30 and 80\,mas\,yr$^{-1}$, exceedingly unlikely based on the kinematic model used in the Besan\c{c}on simulation \citep{Robin:2003jk} described previously. The contrast curves in Fig.~\ref{fig:literature-contrast} also allow us to exclude background galaxies given their negligible proper motion.

The probability of a chance alignment with a unassociated foreground or background brown dwarf is significantly lower than for stellar contaminants. The companion has an SED consistent with brown dwarfs at the L--T transition (Figs.~\ref{fig:cmd} and \ref{fig:spec-chi2}), and the space density of these objects is from $1.4\times10^{-3}$ to $2.0\times10^{-3}$\,pc$^{-3}$ \citep{Reyle:2010gq}. Using the upper estimate and the volume of a cone extending out from the Earth to a distance of 105\,pc (to account for the range of luminosities for these objects) with an angular radius of $34$\,mas, we calculate the probability of a chance alignment to be $\sim$$6.8\times 10^{-9}$. This does not account for the fact an unassociated brown dwarf would have to share a similar proper motion to AF Lep in order for it to remain at a close projected separation from the star such that it would remain undetected in previous high-contrast imaging observations (Fig.~\ref{fig:literature-contrast}). This further reduces the probability of such a chance alignment.

\section{Conclusions}
\label{sec:discussion}
We have unambiguously discovered a point source in close proximity to the young star AF Lep within both the IFS and IRDIS data sets presented in this study. Despite the lack of a second epoch to measure common proper motion, the sum of evidence strongly suggests that we have identified an associated planetary-mass companion, AF Lep b. The near-infrared photometry is similar to that of other young planetary-mass companions (Sect.~\ref{sec:photo}); both the $K1-K2$ and $J_{\rm IFS}-K1$ colors are inconsistent with a distant background object (Fig.~\ref{fig:cmd}), even after accounting for the total column of Galactic dust in the direction of AF Lep. The $JH$ spectroscopy, being most similar to an L--T transition object rather than a distant M star, also strongly favors the companion hypothesis (Figs.~\ref{fig:spec-chi2} and \ref{fig:spec-compare}). Additionally, the companion was imaged at the position predicted from the proper motion anomaly (Fig.~\ref{fig:prediction}) and has a dynamical mass consistent with that estimated from evolutionary models. As discussed in Sect.~\ref{sec:alternatives}, the probability of a chance alignment with an unassociated foreground or background object is very small. The probability of finding one at the position predicted from the proper motion anomaly of AF Lep is smaller still. A measurement of the relative parallactic motion (or lack thereof) between AF Lep and AF Lep b in the coming months will provide further evidence that the companion is bound, although this determination can only be made conclusively after several years of astrometric monitoring.

With a model-independent mass measurement of $4.3_{-1.2}^{+2.9}$\,$M_{\rm Jup}$, AF Lep b is intermediate to the imaged companions \object{$\beta$~Pic~b} and \object{$\beta$~Pic~c} (11.9\,$M_{\rm Jup}$, 8.9\,$M_{\rm Jup}$; \citealp{Lacour:2021cl}) and \object{51 Eri b} (2\,$M_{\rm Jup}$; \citealp{Macintosh:2015ew}). Together with the more massive companions \object{PZ Tel B} (38--72\,$M_{\rm Jup}$; \citealp{Maire:2016go}) and \object{HR 7329 B} ($\eta$~Tel~B, 20--50\,$M_{\rm Jup}$; \citealp{Neuhauser:2011it}), these companions form an isochronal sequence spanning one-and-a-half decades in mass around host stars that assumedly were coeval and therefore have similar abundances. This sequence also spans a large range of effective temperatures: from the late-M (HR 7329 B) and early L-type objects (e.g., $\beta$~Pic~b), through AF Lep b near the L--T transition, to the mid-T 51 Eri b. While the distribution as a function of mass and luminosity is somewhat continuous across the planet--brown dwarf boundary, the orbital separations of these companions may offer clues as to their formation.  The relatively small semimajor axes of the $\beta$ Pic moving group planets ($<15$\,$M_{\rm Jup}$), especially compared to the larger semimajor axes of the two brown dwarf companions, is consistent with these four lower-mass objects forming more like planets than binary stars.  However, a larger demographics analysis would be necessary to definitively identify a breakpoint in formation scenarios.  In addition, future work aimed at uniformly deriving the masses, luminosities, and abundances of these companions will aid in exploring these formation scenarios.

The host star AF Lep (F8V, $1.29\pm0.20$\,$M_{\odot}$) is also less massive than the stellar hosts of most directly imaged planets ($\lesssim 1.6$\,$M_{\odot}$), though still more massive than the Sun.  While wide-separation giant planets are more common around higher-mass stars \citep{Nielsen:2019cb}, the precise form of the stellar mass dependence is not yet clear.  Objects like AF Lep b will shed light on the role stellar mass plays on giant planet formation, which will be aided by a more precise mass for the host star.  Future orbital monitoring of AF Lep b will likely provide a precise dynamical mass for AF Lep.

The measurements and analysis presented in this work only provide limited information regarding the bulk and atmospheric properties of AF Lep b. A more detailed analysis incorporating spectro-photometric follow-up observations spanning a wider wavelength range will be required in order to make reliable measurements of the bolometric luminosity, effective temperature and surface gravity, and composition. In particular, a more optimized observing sequence with SPHERE with a greater amount of field rotation will reduce the effects of residual speckles still present after PSF subtraction. Additionally, continued astrometric monitoring and future {\it Gaia} data releases will refine the dynamical mass measurement, providing an important anchor for evolutionary models given the well-determined age of the host star. There is some urgency with these follow-up observations; 54\% of orbits consistent with the absolute and relative astrometry have the companion at a smaller angular separation at the end of 2023, making it more challenging for high-contrast imaging instruments to detect. The discovery of this companion should also motivate continued efforts to spatially resolve the debris disk exterior to AF Lep b to look for evidence of ongoing planet-disk interactions.

\begin{acknowledgements}
The authors thank the referee for their comments and suggestions that helped to improve the quality of this manuscript. Supported by NASA grant 80NSSC21K0958 (E.L.N and A.E.P). P.K. thanks support from HST-AR-17059 provided by NASA through a grant from STScI under NASA contract NAS5-26555. This research has made use of the SIMBAD database, operated at CDS, Strasbourg, France. This work has made use of data from the European Space Agency (ESA) mission {\it Gaia} (\url{https://www.cosmos.esa.int/gaia}), processed by the {\it Gaia} Data Processing and Analysis Consortium (DPAC, \url{https://www.cosmos.esa.int/web/gaia/dpac/consortium}). Funding for the DPAC has been provided by national institutions, in particular the institutions participating in the {\it Gaia} Multilateral Agreement. Based on observations collected at the European Southern Observatory under ESO programme 109.23AQ.001.  Based on observations obtained with the Apache Point Observatory 3.5-meter telescope, which is owned and operated by the Astrophysical Research Consortium.  ARCES data reduction pipeline originally created by Julie Thorburn, Karen Kinemuchi, and Jean McKeever.  We thank Hannah Gallamore, Adam Smith, and Jessica Klusmeyer for obtaining some of the ARCES spectra used in this work. This research has made use of data reprocessed as part of the ALICE program, which was supported by NASA through grants HST-AR-12652 (PI: R. Soummer), HST-GO-11136 (PI: D. Golimowski), HST-GO-13855 (PI: E. Choquet), HST-GO-13331 (PI: L. Pueyo), and STScI Director's Discretionary Research funds, and was conducted at STScI which is operated by AURA under NASA contract NAS5-26555.
\end{acknowledgements}

\bibliographystyle{aa}
\bibliography{article.bib}

\begin{appendix}

\section{HST Observations}
\label{ap:hst}
AF Lep was coronagraphically imaged twice with the HST. GO-7226 (PI Becklin) was a Near Infrared Camera and Multi-Object Spectrometer (NICMOS) F160W survey for exoplanets around young, nearby stars and GO-10487 (PI Ardila) used the Advanced Camera for Surveys (ACS) High Resolution Channel (HRC) to detect debris disks around $\beta$ Pic moving group stars with infrared excesses. Even though no detections from these observations have been reported in the literature, we reexamine the data to double-check for any evidence of companions or dust scattered light oriented near the $\sim$70$\degr$ position angle of AF Lep b.

The NICMOS observations were made on UT1998-11-14 with AF Lep placed behind the $0\farcs6$ diameter (0.075"/pix) occulting spot and the telescope rolled by 29.9$\degr$ within a single orbit \citep{Lowrance:2005ci}. The data were processed as part of the Archival Legacy Investigations of Circumstellar Environments (ALICE) program \citep{Choquet:2014ej, Hagan:2018ij} that used a PSF library assembled from all similar NICMOS coronagraphic observations to subtract AF Lep's stellar PSF. The high level science products were downloaded from the ALICE archive\footnote{\url{https://archive.stsci.edu/prepds/alice/}} and used for Figs.~\ref{fig:hst_alice} and~\ref{fig:literature-contrast}. No scattered light from the circumstellar disk is evident in Fig.~\ref{fig:hst_alice}.

\begin{figure}[h!]
    \centering
    \includegraphics[width=0.5\textwidth]{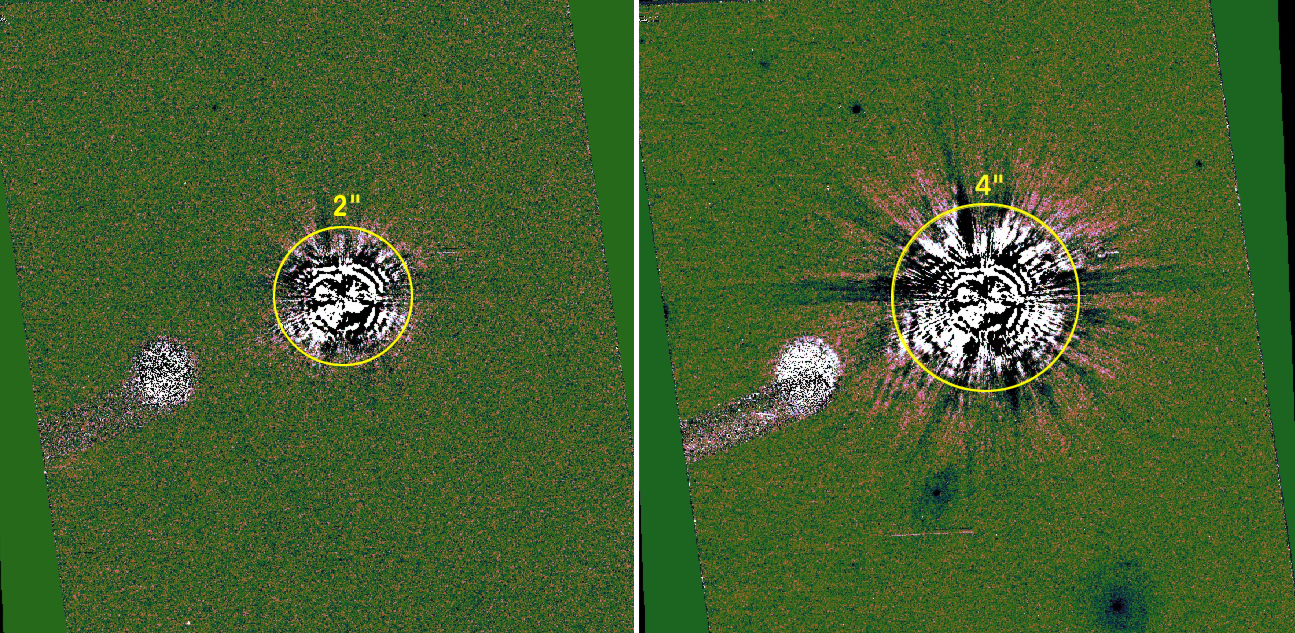}
    \caption{\label{fig:hst_acs}PSF-subtracted F606W HST/ACS/HRC images with the short-exposure data on the left and the long exposure data on the right.  North is up, and east is to the left. The annotations indicate radius in arcseconds. The residual light beyond $4\arcsec$ radius in the long exposure data indicate a poor spectral match between AF Lep and HD 36379. No astrophysical sources are detected in these data.}
\end{figure}

AF Lep was observed on UT2005-09-24 with the HRC ($0\farcs025$/pix) $1\farcs8$ diameter occulting spot in the F606W filter. Reference differential imaging was adopted as the PSF-subtraction strategy (i.e., AF Lep was observed in a single orbit at a single telescope roll orientation) and HD 36379 was observed as the PSF reference star in one orbit immediately after the AF Lep observations. The AF Lep data included one short integration (100 sec, CR-SPLIT=2) and one long integration (2000 sec, CR-SPLIT=4).  Working with the calibrated $\_drz.fits$ files downloaded from the Mikulski Archive, we subtracted the short and long exposure AF Lep data from the corresponding HD 36379 data, iteratively shifting the position and scaling the intensity of the HD 36379 image in order to minimize the PSF subtraction residuals beyond the edge of the occulting spot. Figure~\ref{fig:hst_acs} shows significant residual light in the long exposure data beyond $\sim$$4\arcsec$ radius that is likely due to the spectral mismatch between the stellar PSFs of AF Lep (F8V; $B-V=0.54$) and HD 36379 (G2V; $B-V=0.56$). No other stars were observed in GO-10487 that had an equal or better color match to AF Lep. In general, a detection of dust-scattered light with the ACS/HRC is not expected because the fractional infrared luminosity of AF Lep is roughly four times smaller than the faintest disks that were detected with the instrument \citep{Golimowski:2011dh}.

\clearpage
\section{BKA fit residuals}
\label{ap:bka}
\begin{figure}[h!]
    \centering
    \includegraphics[width=0.5\textwidth]{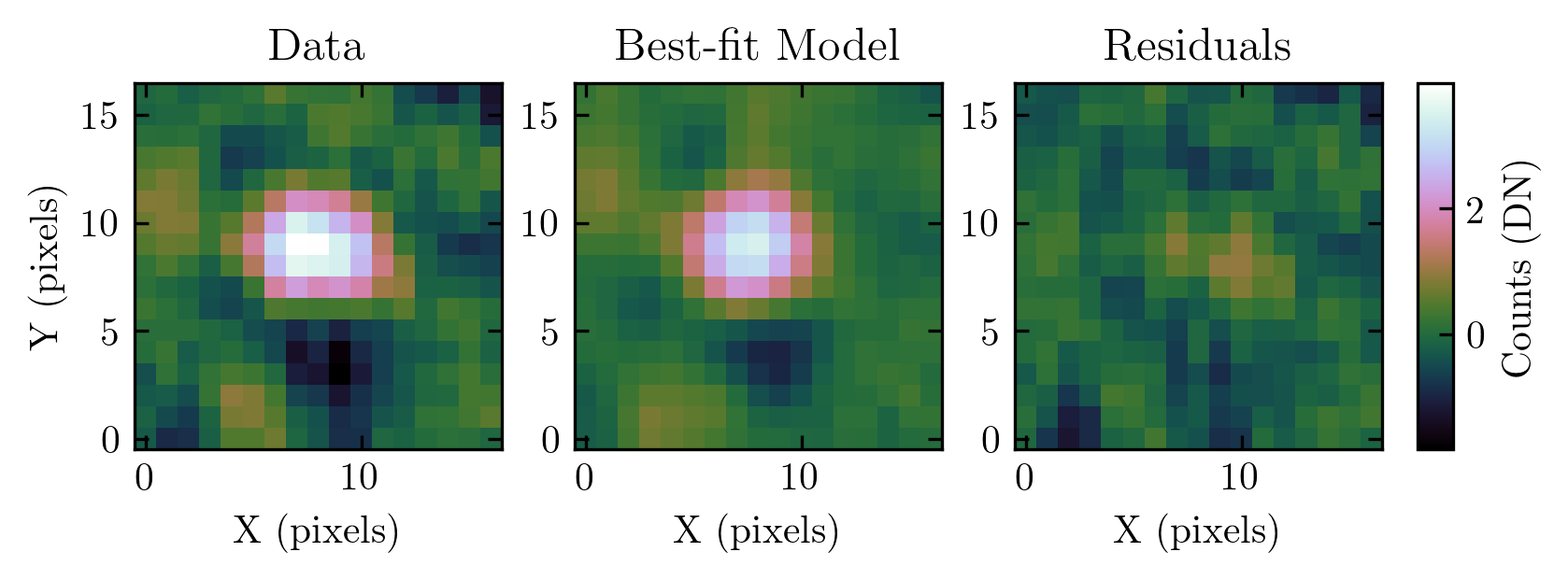}
    \includegraphics[width=0.5\textwidth]{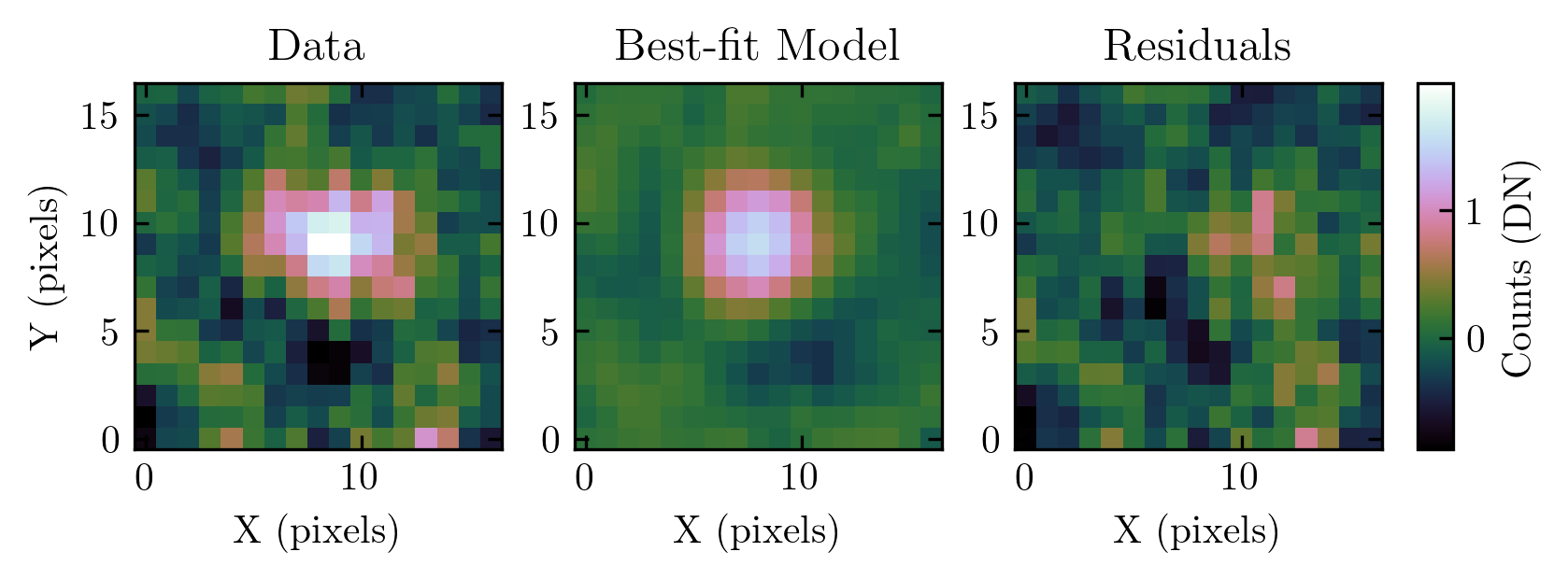}
    \includegraphics[width=0.5\textwidth]{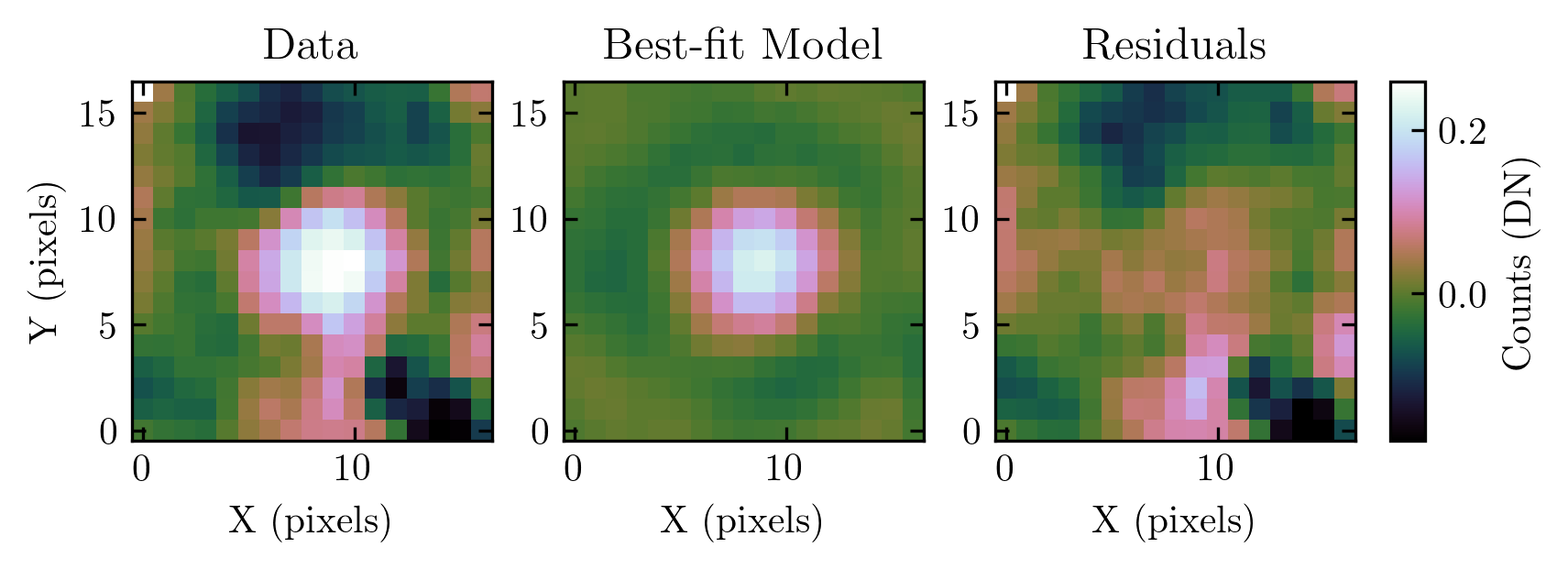}
    \caption{\label{fig:bka}Data (left) best-fit model (middle) and residuals (right) for the BKA fit to AF Lep b in the IRDIS $K1$ (top) and $K2$ (middle) data sets, and the IFS $H2$ data set (bottom).}
\end{figure}

The astrometric and photometric measurements presented in this work were made by fitting a forward-modeled PSF to the PSF-subtracted, temporally (and spectrally) averaged data cubes. The data, best fitting model, and the corresponding residuals are shown in Fig.~\ref{fig:bka} for the two IRDIS data sets, and the $H2$ IFS data set synthesized from the three channels corresponding to the IRDIS $H2$ filter. The best fit models for the two IRDIS data sets are a good match to the data, with little structure remaining in the residual image.

\begin{figure}[h!]
    \centering
    \includegraphics[width=0.5\textwidth]{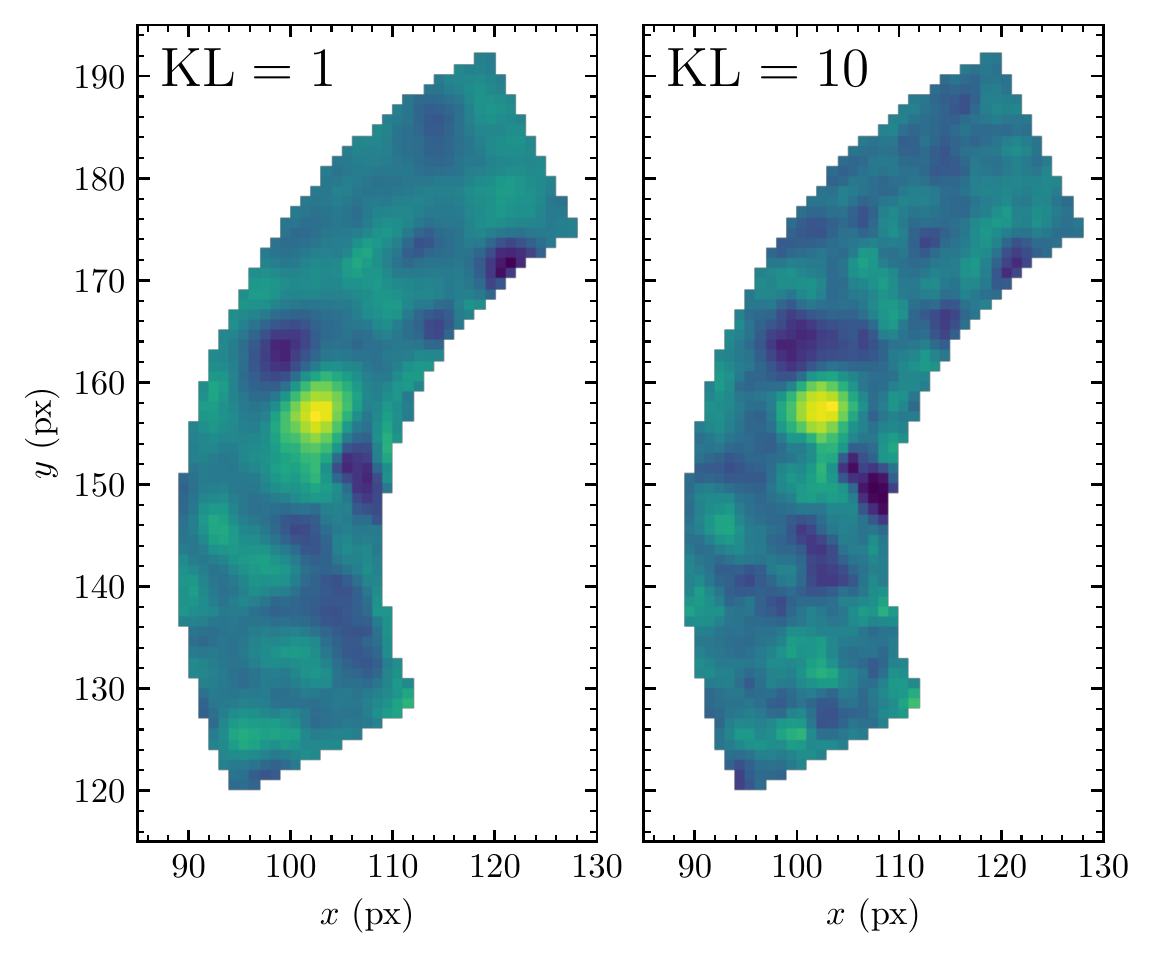}
    \caption{\label{fig:bka-ifs-seg}IFS $H2$ PSF-subtracted segments using a model constructed with one KL basis vector (left) and with ten (right).}
\end{figure}

The best fit for the IFS data set is considerably worse due to the lower S/N of the detection and corresponding large amplitude (both positive and negative) speckles (Fig.~\ref{fig:bka}, bottom panel). The residual noise is significantly structured in the vicinity of AF Lep b, most notably the strong negative residuals to the upper left and bottom right, as shown in Fig.~\ref{fig:bka-ifs-seg}. Between the first ($1.579$\,$\mu$m) and the last ($1.614$\,$\mu$m) wavelength channel the companion appears to move slightly outward. This is consistent with the negative speckle moving radially outward, causing the photocenter of the companion to be shifted. This may be the cause of the apparent discrepancy between the IRDIS and IFS astrometry. Further observations that are timed to maximize field rotation will greatly improve the quality of the PSF subtraction, leading to more reliable astrometric and photometric measurements at these shorter wavelengths.

\clearpage
\onecolumn
\section{Orbital parameter posterior distributions}
\begin{figure}[h!]
    \centering
    \includegraphics[width=0.75\textwidth]{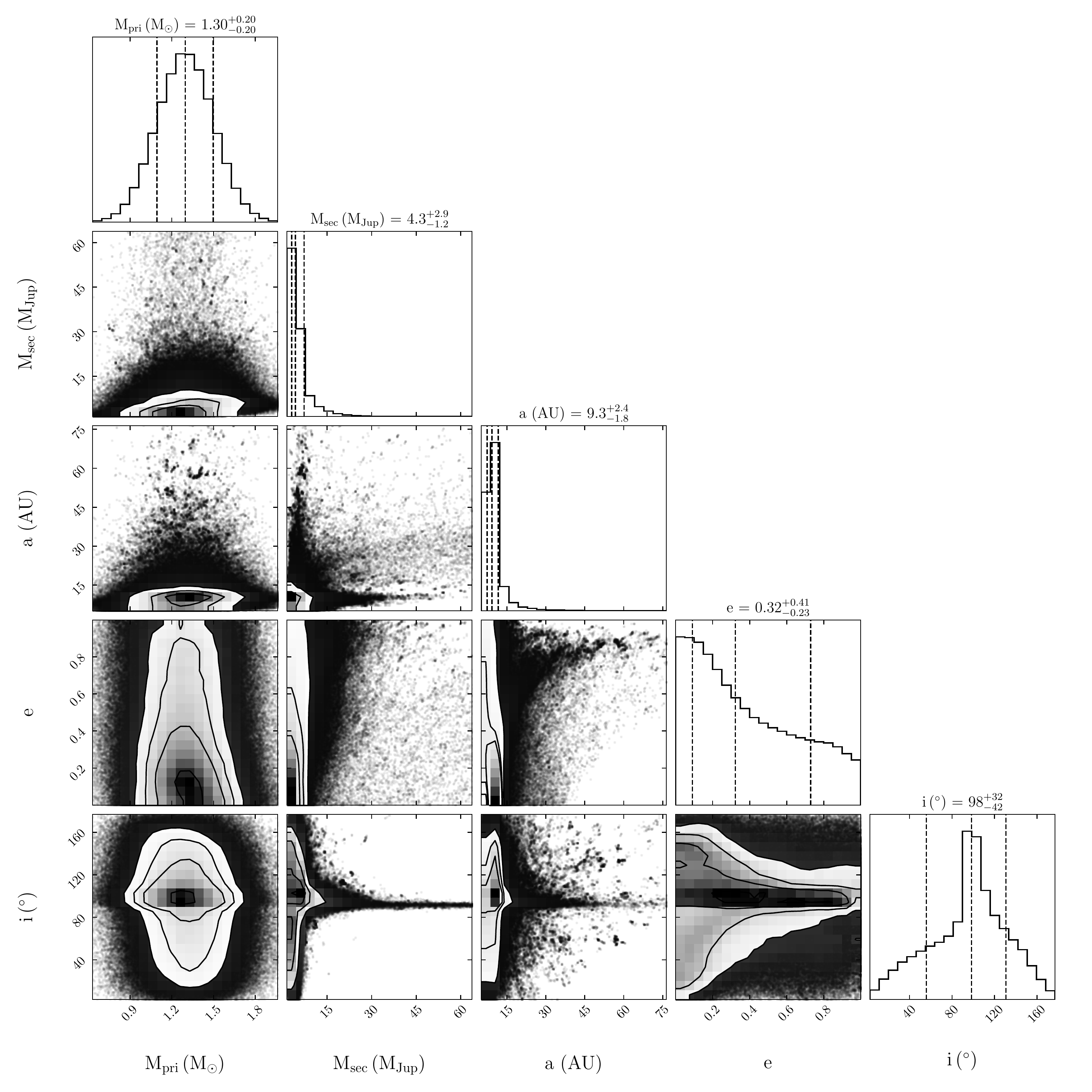}
    \caption{\label{fig:orvara-pdf}Posterior distributions for three of the orbital parameters and the masses of the two components from the {\texttt orvara} fit to the absolute and relative astrometry.}
\end{figure}

Figure~\ref{fig:orvara-pdf} shows the posterior distribution for three of the key orbital parameters, as well for the mass of the primary star and the planetary companion.

\clearpage
\onecolumn
\section{Spectral extraction residuals}
\label{ap:spec-res}
\begin{figure}[h!]
    \centering
    \includegraphics[width=0.5\textwidth]{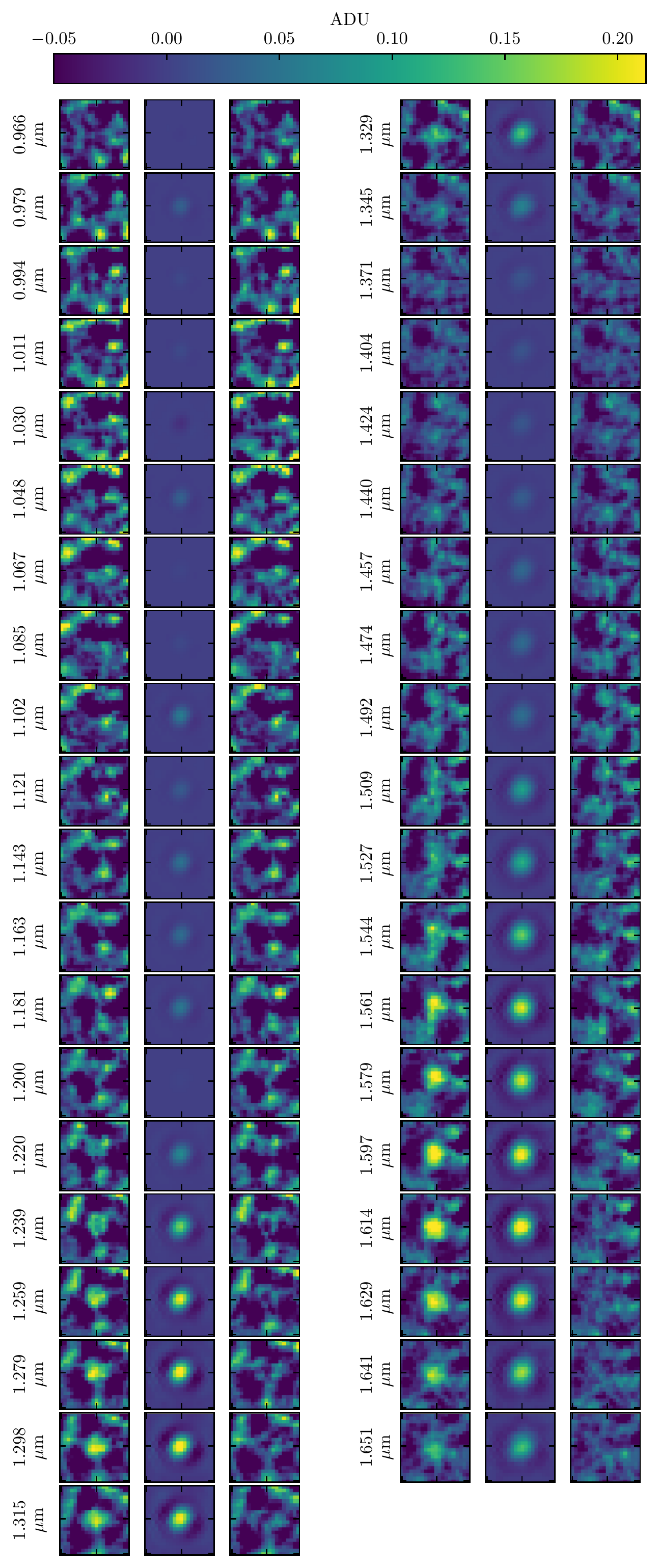}
    \caption{\label{fig:spec-res}Data (left) forward model (middle) and residual (right) for each channel of the PSF-subtracted IFS datacube. The least-squares fit for the bluer channels results in a negative flux and are therefore not plotted.}
\end{figure}
Figure~\ref{fig:spec-res} shows the data, best-fitting forward model, and the residuals for each of the 39 wavelength channels from the spectral extraction analysis described in Sect.~\ref{sec:spectroscopy}.

\end{appendix}

\end{document}